\documentclass{SciPost}

\hypersetup{
    colorlinks,
    linkcolor={red!50!black},
    citecolor={blue!50!black},
    urlcolor={blue!80!black}
}

\usepackage{orcidlink}

\usepackage[bitstream-charter]{mathdesign}
\DeclareSymbolFont{usualmathcal}{OMS}{cmsy}{m}{n}
\DeclareSymbolFontAlphabet{\mathcal}{usualmathcal}

\fancypagestyle{SPstyle}{
\fancyhf{}
\lhead{\colorbox{scipostblue}{\bf \color{white} ~SciPost Physics Lecture Notes }}
\rhead{{\bf \color{scipostdeepblue} ~Submission }}

\fancyfoot[C]{\textbf{\thepage}}
}

\usepackage{braket}
\usepackage{placeins}
\usepackage[british]{babel}

\definecolor{rawGate}     {RGB}{136, 141, 186}   
\definecolor{avgGate}     {RGB}{192, 150, 219}   
\definecolor{gramBond}    {RGB}{224, 202, 240}   
\definecolor{wgNode}      {RGB}{188,  54, 118}   
\definecolor{topBdry}     {RGB}{255, 102,   0}   
\definecolor{initBdry}    {RGB}{240, 188,  66}   
\definecolor{sigmaCircle} {RGB}{255, 204, 153}   
\definecolor{mpsSite}     {RGB}{ 66, 135, 245}   

\usepackage{algorithmicx}
\usepackage{algpseudocode}
\usepackage{amsthm}
\usepackage{mathtools}
\usepackage{tikz}
\usetikzlibrary{patterns,decorations.pathreplacing,decorations.pathmorphing,shapes,arrows.meta,positioning,fit,calc}
\usepackage{booktabs}
\usepackage{tcolorbox}
\tcbuselibrary{listings,skins,breakable}
\usepackage{listings}
\usepackage{hyperref}

\newtcolorbox{funcbox}[1]{%
  enhanced, breakable,
  colback=scipostblue!3!white,
  colframe=scipostblue,
  arc=1.5mm, outer arc=1.5mm,
  boxrule=0.6pt, left=3mm, right=3mm, top=2mm, bottom=2mm,
  title={\ttfamily #1},
  coltitle=white,
  fonttitle=\bfseries\small,
  attach boxed title to top left={xshift=4mm, yshift=-2.2mm},
  boxed title style={colback=scipostblue, colframe=scipostblue,
                     arc=1mm, outer arc=1mm, boxrule=0pt,
                     left=2mm, right=2mm, top=0.5mm, bottom=0.5mm},
  before skip=8pt, after skip=8pt,
}

\newcommand{\rrangle}{\rangle\!\rangle}
\newcommand{\llangle}{\langle\!\langle}
\newcommand{\kket}[1]{|#1\rrangle}
\newcommand{\bbra}[1]{\llangle #1|}
\newcommand{\bbrakket}[2]{\llangle #1|#2\rrangle}
\DeclareMathOperator{\Tr}{Tr}

\definecolor{exerciseBg}    {RGB}{240, 240, 240}   
\definecolor{exerciseFrame} {RGB}{ 80,  80,  88}   
\newcounter{exercisectr}
\renewcommand{\theexercisectr}{\arabic{exercisectr}}
\newenvironment{exercise}{%
  \refstepcounter{exercisectr}%
  \par\medskip
  \begin{tcolorbox}[
    enhanced, breakable,
    colback=exerciseBg,
    colframe=exerciseFrame,
    coltitle=white,
    boxrule=0.6pt, arc=2pt,
    left=8pt, right=8pt, top=5pt, bottom=5pt,
    title={\sffamily\bfseries Exercise~\theexercisectr},
    colbacktitle=exerciseFrame,
  ]%
}{%
  \hfill\textcolor{exerciseFrame}{$\blacksquare$}%
  \end{tcolorbox}\par\medskip%
}

\newcounter{algctr}
\renewcommand{\thealgctr}{\arabic{algctr}}

\newenvironment{Algorithm}[1]{%
  \refstepcounter{algctr}%
  \par\medskip
  \noindent\rule{\linewidth}{0.4pt}\par
  \noindent\textbf{Algorithm~\thealgctr. #1}\par
  \noindent\rule{\linewidth}{0.4pt}\par
  \vspace{0.em}%
  \begin{algorithmic}%
}{%
  \end{algorithmic}%
  \vspace{0.em}%
  \noindent\rule{\linewidth}{0.4pt}\par
  \medskip
}

\definecolor{juliaBg}       {HTML}{F7F7F9}   
\definecolor{juliaFrame}    {HTML}{D8D8DC}   
\definecolor{juliaText}     {HTML}{2A2A32}   
\definecolor{juliaKw}       {HTML}{5E2CA5}   
\definecolor{juliaFn}       {HTML}{1A4FA3}   
\definecolor{juliaType}     {HTML}{8A5A00}   
\definecolor{juliaStr}      {HTML}{996E00}   
\definecolor{juliaComment}  {HTML}{606878}   
\definecolor{juliaMacro}    {HTML}{006E7A}   

\lstdefinelanguage{JuliaCustom}{
    keywords={function, return, end, where, for, if, else, elseif, break, using, begin, struct, mutable, abstract, type, module, import, in, do, let, true, false, nothing},
    morekeywords=[2]{true, false},
    sensitive=true,
    morecomment=[l]{\#},
    morestring=[b]",
    alsoletter={!},
}

\newtcblisting{juliabox}{
    enhanced,
    colback=juliaBg,
    colframe=juliaFrame,
    boxrule=0.6pt,
    arc=1.5mm,
    outer arc=1.5mm,
    top=0mm,
    bottom=0mm,
    left=2mm,
    right=2mm,
    listing only,
    listing engine=listings,
    listing options={language=JuliaCustom},
    before skip=\baselineskip,
    after skip=\baselineskip,
    breakable
}

\lstdefinelanguage{PythonCustom}{
    keywords={def, return, if, else, elif, for, while, in, import, from, as, class, with, try, except, finally, raise, pass, lambda, yield, True, False, None, and, or, not, is, global, nonlocal},
    morekeywords=[2]{True, False, None},
    sensitive=true,
    morecomment=[l]{\#},
    morestring=[b]",
    morestring=[b]',
    alsoletter={\_},
}

\lstdefinelanguage{CppCustom}{
    keywords={int, long, double, float, char, void, bool, if, else, for, while, return, struct, class, namespace, using, template, typename, auto, static, inline, const, public, private, protected, this, true, false, nullptr},
    morekeywords=[2]{Index, ITensor, MPS, SVDResult, Args, replicatn, std, vector, string, size_t},
    sensitive=true,
    morecomment=[l]{//},
    morecomment=[s]{/*}{*/},
    morestring=[b]",
    alsoletter={\_},
}

\newtcblisting{pybox}{
    enhanced,
    colback=juliaBg,
    colframe=juliaFrame,
    boxrule=0.6pt,
    arc=1.5mm,
    outer arc=1.5mm,
    top=0mm,
    bottom=0mm,
    left=2mm,
    right=2mm,
    listing only,
    listing engine=listings,
    listing options={language=PythonCustom},
    before skip=\baselineskip,
    after skip=\baselineskip,
    breakable
}

\newtcblisting{cppbox}{
    enhanced,
    colback=juliaBg,
    colframe=juliaFrame,
    boxrule=0.6pt,
    arc=1.5mm,
    outer arc=1.5mm,
    top=0mm,
    bottom=0mm,
    left=2mm,
    right=2mm,
    listing only,
    listing engine=listings,
    listing options={language=CppCustom},
    before skip=\baselineskip,
    after skip=\baselineskip,
    breakable
}

\begin{document}

\pagestyle{SPstyle}

\begin{center}{\Large \textbf{\color{scipostdeepblue}{
Lecture Notes on Replica Tensor Networks for Random Quantum Circuits\\
}}}\end{center}

\begin{center}\textbf{
Xhek Turkeshi~\orcidlink{0000-0003-1093-3771}\textsuperscript{$\star$}
}\end{center}

\begin{center}
{\bf} Institut f\"ur Theoretische Physik, Universit\"at zu K\"oln, Z\"ulpicher Str.~77, 50937 K\"oln, Germany
\\[\baselineskip]
$\star$ \href{mailto:xturkesh@uni-koeln.de}{\small xturkesh@uni-koeln.de}
\end{center}

\section*{\color{scipostdeepblue}{Abstract}}
\textbf{\boldmath{%
We present a pedagogical, hands-on tutorial on \emph{replica tensor-network} techniques for random quantum circuits. At its core, the method recasts circuit-averaged observables acting on multiple copies of the system as the contraction of a classical tensor network, equivalently the partition function of a statistical-mechanics model whose effective spins live in the commutant of the gate ensemble. The framework is general: changing the observable or the initial state modifies only the replica boundary conditions, while changing the ensemble modifies the bulk tensors. Focusing on quantum-information diagnostics, from metrics of wavefunction spreadings to entanglement quantifiers, we illustrate the approach in both clean and noisy random unitary circuits. We then briefly explain how the methodology extends to other ensembles, such as orthogonal or Clifford circuits. The lecture notes are accompanied by \texttt{ReplicaTN}\textcolor{red}{\footnote{Source code openly available at \url{https://github.com/xturkesh/ReplicaTN}.}}, a self-contained C++/Python library and pedagogical notebooks.
}}

\vspace{\baselineskip}

\noindent\textit{These lecture notes accompany a course I will deliver at the 
\href{https://comp-quant-2026.sciencesconf.org/}{Les Houches School on
Computational Quantum Physics}, June 2026. They are a living
document; corrections, typos, and suggestions are very welcome and
can be sent to the author at
\href{mailto:turkeshi@thp.uni-koeln.de}{\texttt{turkeshi@thp.uni-koeln.de}}.}

\vspace{\baselineskip}

\noindent\textcolor{white!90!black}{%
\fbox{\parbox{0.975\linewidth}{%
\textcolor{white!40!black}{\begin{tabular}{lr}%
  \begin{minipage}{0.6\textwidth}%
    {\small Copyright attribution to authors. \newline
    This work is a submission to SciPost Physics Lecture Notes. \newline
    License information to appear upon publication. \newline
    Publication information to appear upon publication.}
  \end{minipage} & \begin{minipage}{0.4\textwidth}
    {\small Received Date \newline Accepted Date \newline Published Date}%
  \end{minipage}
\end{tabular}}
}}
}

\vspace{10pt}
\noindent\rule{\textwidth}{1pt}
\tableofcontents
\noindent\rule{\textwidth}{1pt}
\vspace{10pt}

\section{Introduction}
\label{sec:introduction}
Random quantum circuits, originally introduced in quantum information and computation, are central in several applications, from certifying quantum computational advantage~\cite{boixo2018characterizing,arute2019quantum} to the theoretical backbone of learning protocols, randomised measurements, and noise characterisation~\cite{Preskill2018quantumcomputingin}.
Today, random circuits also serve as tractable models of generic quantum dynamics~\cite{fisher2023random}. 
Stripped to the essentials, from locality to symmetry, they provide a powerful framework to expose the universal features of far-from-equilibrium quantum matter. 
For instance, in condensed matter theory, they serve as minimal models for thermalisation, entanglement growth, operator spreading~\cite{nahum2017quantum,nahum2018operator,keyserlingk_operator_2018,potter2022entanglement}, and measurement-induced phase transitions~\cite{PhysRevB.109.174307,PhysRevB.101.104302,PhysRevB.101.104301}, while in high-energy physics, they model black-hole scrambling and tackle questions on information recovery~\cite{hayden_black_2007,roberts_chaos_2017,Hayden2016,PhysRevB.102.064202,PhysRevB.100.134203}.
Their success is tied to their tractability: averages over circuit realizations can be recast as classical statistical mechanics problems, whose elementary degrees of freedom are fixed by the gate ensembles. For instance, these spins are permutations for Haar unitary gates~\cite{weingarten_asymptotic_1978,collins_moments_2003,collins_integration_2006,collins_weingarten_2022,mele_introduction_2024}, exact pairings for orthogonal gates~\cite{brauer1937algebras,collins_matsumoto_2009}, and stochastic Lagrangian subspaces for the Clifford ensemble~\cite{gross2021schurweyl,webb_clifford_2016,zhu_multiqubit_2017,nezami,magni2025anticoncentration,bittel_clifford_commutant_2025}.
The resulting object is a classical lattice model in spacetime whose configurations encode how the replicated forward and backward histories are locally paired, glued, or constrained. 
Notably, in certain situations, these models can be resolved exactly or in mean-field approximate regimes~\cite{zhou2019emergent,zhou2}. 
More broadly, averages over circuit realizations of quantum information observables acting on a few copies of the systems, ranging from local purities to collision probabilities,  can be computed efficiently by viewing these statistical mechanics models as the contraction of replica tensor networks. 
To the best of my knowledge, the first numerical implementation comes from Ref.~\cite{rakovszky_diffusive_2019,khemani_operator_2018,keyserlingk_operator_2018} in the context of $U(1)$-symmetric random circuits. 
This was later used in Ref.~\cite{Sahu2024brownian,lovas2024quantum} to study error-resilience or coding transitions~\cite{turkeshi2024errorresilience,dallas2025nonlocalmagicgenerationinformation,dallas2,dallas2026,trigueros2026nonstabilizernesserrorresiliencenoisy,sierant2026theorymagicphasetransitions}, and was systematized in Ref.~\cite{Braccia2024}, where it was extended to other ensembles. 
Later on, this was used in several contexts, from anticoncentration or Hilbert space delocalization in various classes of random circuits~\cite{turkeshi2024hilbert,lami2024anticoncentration,sauliere,magni2025anticoncentration,magni2025anticoncentrationstatedesigndoped,tirrito2025universalspreadingnonstabilizernessquantum,aditya,arman,sauliere2026errorcorrectiontransitionsfinitedepthquantum}, to the spreading of magic resources in chaotic circuits~\cite{turkeshi2025magic,magni2025doped,sierant2026fermionic}. 

This quantum-to-classical replica tensor-network mapping is the organising principle of these lectures. 
Taking a hands-on approach, we first review the mapping to classical statistical-mechanics models, and then explain how the elementary building blocks, replica gates in the bulk and replica boundary conditions, are constructed in concrete examples.
Our main focus is on unitary Haar circuits, without and with noise: here we will study how average metrics of coherences and information spreading evolve through the system as the circuit depth increases. 
As we shall see, changing the observable or the initial state modifies only the replica boundary conditions. 
Additionally, we will show how changing the ensemble modifies the bulk tensors and hence leads to different statistical mechanics models, but at the level of pseudo-code, only a change in the placeholder takes place.

The lecture notes are paired with a short implementation in the companion \texttt{ReplicaTN} library (C++ backend with Python frontend) and pedagogical, user-friendly notebooks to rapidly introduce the reader to state-of-the-art implementations.

\subsection{Outline}
Throughout these notes, we will focus on one-dimensional brickwork circuits of random unitaries from a given ensemble $\mathcal{E}$.
In Sec.~\ref{sec:rtn_setup}, we present how, in individual realizations of the circuit, different quantum information metrics that are non-linear in the state, from the inverse participation ratios to the local purity, can be recast formally as transition amplitudes in $k$-copies, or replica, space for a suitable $k$. 

Averaging over these amplitudes, we recast the replica tensor network, built from the averages of individual gates, whose expression is obtained through Schur-Weyl duality and whose contractions go by the name of \textit{Weingarten calculus}. 
For Haar-random unitaries on $k$-replicas, these gates describe interactions on $k!$-dimensional spins associated to permutations in $\mathrm{S}_k$. 
We explain how this quantity can be evaluated within tensor network contraction, and why this is much more efficient than doing standard tensor network evolution of individual realizations of the circuit and then averaging them. 
An optional subsection~\S\ref{sec:irrep_reduction} explains how for qubit systems, using the irreducible representations exactly reduces the $k!$-dimensional raw basis to a $\mathrm{C}_k$, with no loss of information and a sizeable gain in numerical efficiency.

In Sec.~\ref {sec:noisy_circuits}, we present how the method is extended when the random circuit is interspersed with non-random operations.  
To showcase the power of the framework, we consider the action of single-qudit quantum noise acting on every qudit throughout the system and study, as concrete examples, the relative entropy of coherence, the coherent information, and the asymmetric linear cross-entropy benchmark (XEB). 

In Sec.~\ref{sec:beyond_haar_unitary}, we lift the replica tensor networks to broader gate ensembles, focusing on orthogonal and Clifford random circuits as concrete examples.

\FloatBarrier
\section{Replica tensor networks for unitary brickworks}
\label{sec:overview}
\label{sec:rtn_setup}

\subsection{Brickwork circuits and observables}
\label{sec:rtn:observables}
We consider a one-dimensional chain of $N$ qudits, each with local Hilbert space $\mathcal{H}_d \cong \mathbb{C}^d$, where $d \geq 2$. The total Hilbert space is $\mathcal{H} = \mathcal{H}_d^{\otimes N}$, with dimension $D = d^N$.  

The dynamics are generated by a brickwork quantum circuit. 
Labeling qudits by $i = 1, \ldots, N$, the circuit consists of a sequence of layers of two-qudit unitary gates applied in a staggered pattern:
\begin{equation}
U^{(2m)} = \prod_{i=1}^{N/2-1} U_{2i,2i+1}, \qquad U^{(2m+1)} = \prod_{i=1}^{N/2} U_{2i-1,2i},
\label{eq:layers}
\end{equation}
where $U_{i,j} \in \mathrm{U}(d^2)$ is a unitary gate acting on qudits $i$ and $j$. 
The circuit of depth $t$ is then the product $\mathbf{U}_t = \prod_{r=1}^{t} U^{(r)}$, and the time-evolved state reads $|\Psi_t\rangle = \mathbf{U}_t |\Psi_0\rangle$.

This architecture has a strict \emph{causal light cone}: information propagates at most one site per time step, giving a maximal velocity $v_{\max} = 1$. This is the discrete-time analogue of the Lieb--Robinson bound for Hamiltonian systems~\cite{lieb1972finite}.

\subsubsection{Random brickwork circuits and the inverse participation ratio}

In a random quantum circuit, each two-qudit gate $U_{i,j}$ is drawn independently from the Haar measure on $\mathrm{U}(d^2)$.  We denote a realization of this circuit graphically by
\begin{equation}
\mathbf{U}_t |\Psi_0\rangle
\;=\;
\begin{tikzpicture}[baseline=(current bounding box.center),scale=0.6]
\definecolor{skyromeblue}{RGB}{86,180,233}
\def\nqubits{6}
\def\ndepth{5}
\foreach \x in {1,...,\nqubits}{
    \draw[thick] (\x,0) -- (\x,\ndepth+0.5);
    \draw[thick, fill=white] (\x,0) circle (0.2);
    \node[scale=0.75] at (\x,-0.6) {\large $|0\rangle$};
}
\foreach \d in {0,...,\numexpr\ndepth-1\relax} {
    \pgfmathsetmacro{\y}{\d + 1}
    \ifodd\d
        \foreach \x in {2,4} {
            \draw[thick, draw=skyromeblue!80!black, fill=skyromeblue, rounded corners=2pt]
                (\x-0.2,\y-0.3) rectangle (\x+1.2,\y+0.3);
        }
    \else
        \foreach \x in {1,3,5} {
            \draw[thick, draw=skyromeblue!80!black, fill=skyromeblue, rounded corners=2pt]
                (\x-0.2,\y-0.3) rectangle (\x+1.2,\y+0.3);
        }
    \fi
}
\end{tikzpicture}\;,
\label{eq:rmpsBrickwall}
\end{equation}
where every blue brick is a fresh Haar-random two-qudit unitary.  We will also need the same brickwork with all gates conjugated, $\mathbf{U}_t^* |\Psi_0\rangle$, which we draw in a slightly lighter shade,
\begin{equation}
\mathbf{U}_t^* |\Psi_0\rangle
\;=\;
\begin{tikzpicture}[baseline=(current bounding box.center),scale=0.6]
\definecolor{lightsteelblue}{RGB}{132,178,220}
\def\nqubits{6}
\def\ndepth{5}
\foreach \x in {1,...,\nqubits}{
    \draw[thick] (\x,0) -- (\x,\ndepth+0.5);
    \draw[thick, fill=white] (\x,0) circle (0.2);
    \node[scale=0.75] at (\x,-0.6) {\large $|0\rangle$};
}
\foreach \d in {0,...,\numexpr\ndepth-1\relax} {
    \pgfmathsetmacro{\y}{\d + 1}
    \ifodd\d
        \foreach \x in {2,4} {
            \draw[thick, draw=lightsteelblue!80!black, fill=lightsteelblue, rounded corners=2pt]
                (\x-0.2,\y-0.3) rectangle (\x+1.2,\y+0.3);
        }
    \else
        \foreach \x in {1,3,5} {
            \draw[thick, draw=lightsteelblue!80!black, fill=lightsteelblue, rounded corners=2pt]
                (\x-0.2,\y-0.3) rectangle (\x+1.2,\y+0.3);
        }
    \fi
}
\end{tikzpicture}\;.
\label{eq:rmpsBrickwall_conj}
\end{equation}
Every gate is therefore a ``maximally random'' unitary, with no preferred basis or direction.  The resulting ensemble of circuits, $\{|\Psi_t\rangle\}$, is a distribution over quantum states whose properties we wish to characterise.

The first non-linear
quantity we will track is the \emph{inverse participation ratio}
(IPR): given a state $|\Psi\rangle$, the $k$th IPR with respect to
the computational basis
$\mathcal{B} = \{|\pmb{x}\rangle\}_{\pmb{x} \in \mathbb{Z}_d^N}$ is the $k$th
moment of the Born distribution,
\begin{equation}
\mathcal{I}^{(k)}(|\Psi\rangle)
   = \sum_{\pmb{x} \in \mathbb{Z}_d^N} p_{\pmb{x}}^{\,k} ,
\qquad
p_{\pmb{x}} = |\langle \pmb{x}|\Psi\rangle|^2 .
\label{eq:ipr_def}
\end{equation}
Normalisation forces $\mathcal{I}^{(1)} = 1$; the $k = 2$ case is the \emph{collision probability} and quantifies how delocalised $|\Psi\rangle$ is over $\mathcal{B}$ (a computational basis state has $\mathcal{I}^{(2)} = 1$, the maximally mixed distribution has $\mathcal{I}^{(2)} = 1/D$).  The IPRs are equivalent to the R\'enyi participation entropies $\mathrm{S}_k = -\log \mathcal{I}^{(k)} / (k - 1)$. 
In Sec.~\ref{sec:rtn:average} we compute the Haar-random expectation of~Eq.~\eqref{eq:ipr_def} explicitly via Weingarten calculus and use it to define the notion of \emph{anticoncentration} of an ensemble of states.

A second non-linear quantity that will recur throughout these notes is the R\'enyi-$k$ \emph{local purity} (LP) of a subsystem $A=\{1,\dots,\ell\}$ in the bipartition $A \cup B = \{1, \ldots, N\}$,
\begin{equation}
\mathcal{P}^{(k)}_A(|\Psi\rangle)  \;=\;  \Tr_A\!\left( \rho_A^{k}\right) ,
\qquad
\rho_A \;=\; \Tr_B \, |\Psi\rangle\!\langle\Psi| ,
\label{eq:purity_def}
\end{equation}
which probes the entanglement between $A$ and its complement.  For $k = 2$, $\mathcal{P}^{(2)}_A = \Tr(\rho_A^2)$ is the standard linear purity, ranging from $1/D_A$ (maximally mixed $\rho_A$) to $1$
(separable pure state on $A\cup B$).  The associated R\'enyi entanglement entropies are $S^{(k)}_A = -\log \mathcal{P}^{(k)}_A / (k - 1)$.  The Haar-random expectation of $\mathcal{P}^{(2)}_A$, the Page value, is also derived in Sec.~\ref{sec:rtn:average} as the second worked example.

\smallskip\noindent
The inverse participation ratios~Eq.~\eqref{eq:ipr_def} and the local purity~Eq.~\eqref{eq:purity_def}
are the prototype examples we will focus in this section: they are non-linear in the state, and their circuit average admits an exact replica representation in the doubled Hilbert space.
The replica orders $k = 2$ and $k = 3$ will recur explicitly in later sections: for instance $k = 2$ controls also the linear cross-entropy benchmark, while $k = 3$ is the smallest replica number that distinguishes the Clifford ensemble from the unitary one for qudits of dimension $d>2$~\cite{webb_clifford_2016,zhu_multiqubit_2017}.

\subsection{Superoperator formalism and replica space}
\label{sec:rtn:superoperator}

To compute circuit averages of any quantum-information metric that is \emph{non-linear} in the state, the workhorse is \emph{Weingarten calculus}: a closed-form expansion for polynomial moments of a Haar-random unitary.  The IPR and the LP of~\S\ref{sec:rtn:observables} are two prototypical examples, but the same machinery extends to relative entropies of coherence, coherent information, frame potentials, linear cross-entropies, and beyond.  
The idea is simple: non-linear observables in the state can be built as a linear functional in multiple copies, or replicas, of the system. 
The dual vector that induces this functional is the only information when the observable of interest $\Omega_k$ enters.
Then the average over this functional, by linearity, is the functional over the average, which can be evaluated through Schur-Weyl duality.
As we shall see, this leads to the partition function of a classical statistical mechanics model.
For convenience, we introduce the doubled-Hilbert-space representation here; the actual averaging is carried out in~\S\ref{sec:rtn:average}.

Every operator $A \in \mathrm{End}(\mathcal{H})$ on a $D$-dimensional
Hilbert space $\mathcal{H}$ can be vectorised as a state
$|A\rrangle \in \mathcal{H} \otimes \mathcal{H}$ via the
Choi--Jamio{\l}kowski isomorphism.  Concretely, one introduces the
unnormalised maximally entangled vector
\begin{equation}
|\varphi\rangle \;=\; \sum_{\pmb{x} \in \mathbb{Z}_d^N}\,|\pmb{x}, \pmb{x}\rangle ,
\label{eq:choi_phi}
\end{equation}
and define the doubled vector by acting on the second copy:
\begin{equation}
|A\rrangle \;\equiv\; (\mathbb{I} \otimes A)\,|\varphi\rangle
\;=\; \sum_{\pmb{x}, \pmb{y} \in \mathbb{Z}_d^N}\, A_{\pmb{y}\pmb{x}}\,|\pmb{x}, \pmb{y}\rangle .
\label{eq:choi_def}
\end{equation}
This convention immediately reproduces the three identities we will
use throughout:
\begin{equation}
A \mapsto |A\rrangle , \qquad
\mathrm{tr}(A^\dagger B) = \bbrakket{A}{B} , \qquad
U A U^\dagger \mapsto (U \otimes U^*)\,|A\rrangle ,
\label{eq:choi_props}
\end{equation}
where the inner-product identity is direct verification:
$\bbrakket{A}{B} = \sum_{\pmb{x}, \pmb{y}} A^*_{\pmb{y}\pmb{x}} B_{\pmb{y}\pmb{x}} = \mathrm{tr}(A^\dagger B)$,
using the unnormalised $|\varphi\rangle$ of~Eq.~\eqref{eq:choi_phi}.
The third identity, that conjugation $A \mapsto U A U^\dagger$
becomes the simple linear action $(U \otimes U^*)$ on the doubled
vector, is the key reason for switching to this representation: for a given initial state  $\rho_0=|\Psi_0\rangle\langle \Psi_0|$,
the time evolution of any observable expectation
$\mathrm{tr}(\Omega\,\rho_t)$ collapses to a plain transition amplitude
$\llangle \Omega | (\mathbf{U}_t \otimes \mathbf{U}_t^*)|\rho_0\rrangle$
between two vectors in $\mathcal{H} \otimes \mathcal{H}$.

For a degree-$k$ polynomial observable in the state amplitudes, the average over
the random circuit produces a closed expression only on the
\emph{$k$-replica space} $\mathcal{H}^{\otimes 2k}$, obtained by
taking $k$ statistically identical copies of the doubled-space
evolution.  We introduce two pieces of notation that we will use
throughout the rest of the manuscript.

Stacking $k$ statistically identical
copies of the doubled-space evolution $\mathbf{U}_t \otimes
\mathbf{U}_t^*$ produces a \emph{replica operator} acting on the
$2k$-replica Hilbert space $\mathcal{H}^{\otimes 2k} \;\equiv\;
(\mathcal{H} \otimes \mathcal{H})^{\otimes k}$,
\begin{equation}
\bigl(\mathbf{U}_t \otimes \mathbf{U}_t^*\bigr)^{\otimes k}
   \;:\;
   \mathcal{H}^{\otimes 2k} \;\to\; \mathcal{H}^{\otimes 2k} ,
\label{eq:replica_op}
\end{equation}
and the corresponding $k$-replicated reference vector is
$|\rho_0^{\otimes k}\rrangle \in \mathcal{H}^{\otimes 2k}$.  We label
the replicas by an index $m = 1, \ldots, k$, and write the
basis of $(\mathcal{H} \otimes \mathcal{H})^{\otimes k}$ as
$|{\pmb{b}}_1, \bar {\pmb{b}}_1, \ldots, {\pmb{b}}_k, \bar {\pmb{b}}_k\rangle$, where ${\pmb{b}}_m$ and
$\bar {\pmb{b}}_m \in \mathbb{Z}_d^N$ are the ``ket'' and ``bra''
indices of replica $m$.

On the $k$-replica space, the symmetric
group $\mathrm{S}_k$ acts by permuting the replicas.  Each
$\sigma \in \mathrm{S}_k$ defines a \emph{permutation state}, or
\emph{replica spin},
$|\sigma\rrangle \in \mathcal{H}^{\otimes 2k}$ via
\begin{equation}
\langle \pmb{b}_1, \bar {\pmb{b}}_1, \ldots, \pmb{b}_k, \bar {\pmb{b}}_k\,|\,\sigma\rrangle
\;=\;
\prod_{m = 1}^{k}\,\delta_{{\pmb{b}}_m,\,\bar {\pmb{b}}_{\sigma(m)}} ,
\label{eq:perm_state}
\end{equation}
which is the doubled-space image of the operator on
$\mathcal{H}^{\otimes k}$ that permutes the $k$ replica copies
according to $\sigma$.  At $k = 2$, $\mathrm{S}_2 = \{\iota, s\}$ contains only identity and swap, and the corresponding permutation states are
\begin{equation}
|\iota\rrangle \;=\; \sum_{\pmb{b}_1, \pmb{b}_2 \in \mathbb{Z}_d^N}|\pmb{b}_1, \pmb{b}_1, \pmb{b}_2, \pmb{b}_2\rangle ,
\qquad
|s\rrangle \;=\; \sum_{\pmb{b}_1, \pmb{b}_2 \in \mathbb{Z}_d^N}|\pmb{b}_1, \pmb{b}_2, \pmb{b}_2, \pmb{b}_1\rangle ,
\label{eq:iota_swap_states}
\end{equation}
the first being a pair of independent traces on the two replicas
and the second a swap of the two replicas.  
As we shall see, the permutation states will play the role of
``classical spins'' in the averaged tensor network of
\S\ref{sec:rtn:average}. They are not orthogonal: their inner product is the
\emph{Gram entry}
\begin{equation}
G_{\sigma, \pi}(q) \;=\; \llangle \sigma\,|\,\pi\rrangle
\;=\; q^{\#(\sigma^{-1}\pi)},
\label{eq:gram_def}
\end{equation}
where $\#(\tau)$ is the number of cycles of the permutation $\tau$
and $q$ is the dimension of the single-replica space, e.g.\ $q = d$
on a single site, $q = D$ for the whole system. 

The inverse participation ratios, cf. Eq.~\eqref{eq:ipr_def}, and the local purity, cf. Eq.~\eqref{eq:purity_def}, fit into a single common template: each of them is the doubled-replica-space
overlap
\begin{equation}
\mathbb{E}\bigl[\,\Lambda(|\Psi_t\rangle)\,\bigr]
\;=\;
\llangle \hat\Omega_k^\Lambda \, \big|\,
   \hat{\Phi}^{(k)}_t\,
   \big| \hat{\mathfrak{I}}_k\rrangle ,
\label{eq:generic_overlap}
\end{equation}
where $|\hat{\mathfrak{I}}_k\rrangle:=|\rho_0^{\otimes k}\rrangle$ is the $k$-replica initial state, $\hat{\Phi}_t^{(k)}= \mathbb{E}\bigl[(\mathbf{U}_t \otimes \mathbf{U}_t^*)^{\otimes k}\bigr]$ is the so-called $k$th \textit{twirling channel}, and $|\hat\Omega_k^\Lambda\rrangle$ is the observable-dependent boundary ket.  Only the boundary changes between observables; the bulk averaging is the same.  
We keep $|\hat{\Omega}_k^\Lambda\rrangle$ generic whenever possible, and now write
down the two specific boundaries we will use throughout these notes.

Substituting the IPR~Eq.~\eqref{eq:ipr_def} into the
template~Eq.~\eqref{eq:generic_overlap} gives
\begin{equation}
\mathcal{I}^{(k)}(|\Psi_t\rangle)
\;=\;
\sum_{\pmb{x}}\,\mathrm{tr}\bigl(|\pmb{x}\rangle\langle\pmb{x}|\cdot|\Psi_t\rangle\langle\Psi_t|\bigr)^k
\;=\;
\llangle \hat\Omega^{\textup{IPR}}_k | (\mathbf{U}_t \otimes \mathbf{U}_t^*)^{\otimes k} |\hat{\mathfrak{I}}_k\rrangle ,
\label{eq:ipr_replica}
\end{equation}
where the IPR boundary state is
\begin{equation}
|\hat\Omega^{\textup{IPR}}_k\rrangle
\;=\;
\sum_{\pmb{x} \in \mathbb{Z}_d^N}\,\bigl(|\pmb{x},\pmb{x}\rangle\bigr)^{\otimes k} \;.
\label{eq:boundary_IPR}
\end{equation}
This forces all $k$ ket-copies and all $k$ bra-copies to
take the same computational-basis value, projecting the
$k$-replica space onto its diagonal subspace.

Consider a contiguous bipartition $A \cup B$ with $A = \{1, \ldots, \ell\}$ and $B = \{\ell + 1, \ldots, N\}$.  The $k$th local purity $\mathcal{P}^{(k)}_A = \mathrm{tr}(\rho_A^k)$ of the complement region has the same template form
\begin{equation}
\mathcal{P}^{(k)}_A(|\Psi_t\rangle)
\;=\;
\llangle \hat\Omega^{\textup{LP}}_k |
        (\mathbf{U}_t \otimes \mathbf{U}_t^*)^{\otimes k}
        |\hat{\mathfrak{I}}_k\rrangle ,
\label{eq:boundary_LP}
\end{equation}
with the boundary ket 
\begin{equation}
|\hat{\Omega}^{\textup{LP}}_k\rrangle
\;=\sum_{\substack{\pmb{a}_1,\pmb{a}_2,\dots, \pmb{a}_k\in \mathbb{Z}_d^\ell \\ \pmb{b}_1,\pmb{b}_2,\dots, \pmb{b}_k \in \mathbb{Z}_d^{N-\ell}}} |(\pmb{a}_1,\pmb{b}_1),(\pmb{a}_2,\pmb{b}_1),(\pmb{a}_2,\pmb{b}_2),(\pmb{a}_3,\pmb{b}_2)\dots,(\pmb{a}_k,\pmb{b}_k),(\pmb{a}_1,\pmb{b}_k)\rangle
\label{eq:boundary_LP_kets}
\end{equation}
where $(\pmb{a},\pmb{b})$ have $\pmb{a}$ and $\pmb{b}$ running in the Hilbert space of $A$ and $B$, respectively.
We will come back repeatedly to the case $k = 2$, given its relevance for the case study in these notes.
In this regime, $\eta=s$ corresponds to the previously introduced swap operator. 

From the above discussion, it is clear that changing the quantity of interest requires simply changing this boundary operator. In the following subsection, we tackle the ensemble average and show how the replica transfer matrix arises.  

Before taking the average and completing the mapping to a classical statistical mechanics model, it is convenient to reshape, or change the basis, on the replica Hilbert spaces $\frak{A}\colon (\mathcal{H}_d^{\otimes N})^{\otimes k}\to (\mathcal{H}_d^{\otimes k})^{\otimes N}$ in Eq.~\eqref{eq:generic_overlap}. The action of this unitary transformation is given by $|\mathfrak{I}^{(k)}\rrangle:=\mathfrak{A}|\hat{\mathfrak{I}}^{(k)}\rrangle$ and $|\Omega_k^\Lambda\rrangle:=\mathfrak{A}|\hat{\Omega}_k^\Lambda\rrangle$ for initial and final states respectively, and $\Phi_t^{(k)}:=\mathfrak{A}\hat{\Phi}_t^{(k)}\mathfrak{A}^\dagger$ for the bulk operation. Then, it is clear that
\begin{equation}
\mathbb{E}\bigl[\,\Lambda(|\Psi_t\rangle)\,\bigr]
\;=\;\llangle \hat\Omega_k^\Lambda \, \big|\,
   \hat{\Phi}^{(k)}_t\,
   \big| \hat{\mathfrak{I}}^{(k)}\rrangle\;=\;
\llangle \Omega_k^\Lambda \, \big|\,
   \Phi_t^{(k)}\,
   \big| \mathfrak{I}^{(k)}\rrangle\;.
\label{eq:generic_overlap2}
\end{equation}
With this convention, the boundary operators already admit a simpler form.
For instance, the IPR ket $|\Omega^{\textup{IPR}}_k\rrangle=\bigotimes_{i=1}^N|\Omega^{\textup{IPR}}_k\rrangle_i$ factorizes across sites as the \emph{diagonal projector} in the $k$-replica space, with
\begin{equation}
|\Omega^{\textup{IPR}}_k\rrangle_i
\;=\;
\sum_{x_i = 0}^{d - 1}\,\bigl(|x_i, x_i\rangle\bigr)^{\otimes k} \;,
\label{eq:boundary_IPR}
\end{equation}
the boundary operator on each replica qudit.

Also, the boundary operator for the local purity factorised in the replica-spin basis as
\begin{equation}
|\Omega^{\textup{LP}}_k\rrangle
\;=\;
\bigotimes_{i \in A}\,|\eta\rrangle_i
\;\otimes\;
\bigotimes_{i \in B}\,|\iota\rrangle_i ,
\label{eq:boundary_LP_kets}
\end{equation}
where $\iota,\;\eta \in  \mathrm{S}_k$ are $k$-replica state on individual qudits.

We now turn to the bulk average of the circuit, assuming implicitly that the reordering $\mathfrak{A}$ is performed when necessary.

\subsection{Ensemble average}
\label{sec:rtn:average}

We now perform the Haar average of the bulk
operator ${\Phi}_t^{(k)}$. 
Since in the circuit $\mathbf{U}_t$ each gate is identically and randomly distributed according to the unitary ensemble $\mathrm{U}(d^2)$, by statistical independence, the average over the whole circuit is built up from the average of each individual gate.  
The statistical average couples the $k$ replicas at every gate, and the stack collapses onto a \emph{single} classical 2-D tensor network
whose vertical legs carry one $\mathrm{S}_k$-spin per site:
\begin{equation}
\Phi^{(k)}_t \;\equiv\;
\mathbb{E}_\mathrm{Haar}\!\left[
\begin{tikzpicture}[baseline=(current bounding box.center),scale=0.55]
\definecolor{skyromeblue}{RGB}{86,180,233}
\definecolor{lightsteelblue}{RGB}{132,178,220}
\definecolor{darkgraylines}{RGB}{55,55,55}
\def\nqubits{6}
\def\ndepth{5}
\def\repsep{0.24}
\def\gateborder{0.9pt}
\draw[decorate, decoration={brace, amplitude=4pt}, thick, darkgraylines]
  (0.25,0.0) -- node[left=4pt]{\scriptsize $t$} (0.25,\ndepth+0.5);
\draw[decorate, decoration={brace, amplitude=4pt, mirror}, thick, darkgraylines]
  ({6.9},{\ndepth+0.15}) --
  node[right=4pt]{\scriptsize $k$}
  ({6.9+5*\repsep},{\ndepth+0.15+5*\repsep});
\foreach \kk[evaluate=\kk as \xshift using \repsep*\kk,
             evaluate=\kk as \yshift using \repsep*\kk] in {5,4,3,2,1,0}{
    \pgfmathtruncatemacro{\isconj}{mod(\kk,2)}
    \ifnum\isconj=0
        \def\gatecolor{skyromeblue}
    \else
        \def\gatecolor{lightsteelblue}
    \fi
    \foreach \x in {1,...,\nqubits}{
        \draw[thick, darkgraylines]
          (\x+\xshift,0+\yshift) -- (\x+\xshift,\ndepth+0.5+\yshift);
    }
    \foreach \d in {0,...,\numexpr\ndepth-1\relax}{
        \pgfmathsetmacro{\y}{\d + 1}
        \ifodd\d
            \foreach \x in {2,4}{
                \draw[line width=\gateborder, draw=darkgraylines,
                      fill=\gatecolor, rounded corners=2pt]
                  (\x-0.2+\xshift,\y-0.3+\yshift) rectangle
                  (\x+1.2+\xshift,\y+0.3+\yshift);
            }
        \else
            \foreach \x in {1,3,5}{
                \draw[line width=\gateborder, draw=darkgraylines,
                      fill=\gatecolor, rounded corners=2pt]
                  (\x-0.2+\xshift,\y-0.3+\yshift) rectangle
                  (\x+1.2+\xshift,\y+0.3+\yshift);
            }
        \fi
    }
}
\end{tikzpicture}
\right]
\;=\;
\begin{tikzpicture}[baseline=(current bounding box.center),scale=0.55]
\definecolor{crimsonred}{RGB}{66,135,245}    
\definecolor{darkgraylines}{RGB}{55,55,55}
\definecolor{dotorange}{RGB}{245,130,32}
\def\nqubits{6}
\def\ndepth{5}
\def\gateborder{0.9pt}
\def\bulletsize{3.5pt}
\draw[decorate, decoration={brace, amplitude=4pt}, thick, darkgraylines]
  (0.25,0.0) -- node[left=4pt]{\scriptsize $t$} (0.25,\ndepth+0.5);
\foreach \x in {1,...,\nqubits}{
    \draw[thick, darkgraylines] (\x,0) -- (\x,\ndepth+0.5);
}
\foreach \d in {0,...,\numexpr\ndepth-1\relax}{
    \pgfmathsetmacro{\y}{\d + 1}
    \ifodd\d
        \foreach \x in {2,4}{
            \draw[line width=\gateborder, draw=darkgraylines,
                  fill=crimsonred, rounded corners=2pt]
              (\x-0.2,\y-0.3) rectangle (\x+1.2,\y+0.3);
        }
    \else
        \foreach \x in {1,3,5}{
            \draw[line width=\gateborder, draw=darkgraylines,
                  fill=crimsonred, rounded corners=2pt]
              (\x-0.2,\y-0.3) rectangle (\x+1.2,\y+0.3);
        }
    \fi
}
\foreach \x/\y in {%
    1/2.5,2/1.5,3/1.5,4/1.5,5/1.5,6/2.5
    ,2/2.5,3/2.5,4/2.5,5/2.5,
    1/4.5,2/3.5,3/3.5,4/3.5,5/3.5,6/4.5
    ,2/4.5,3/4.5,4/4.5,5/4.5}{
    \draw[line width=0.45pt, draw=darkgraylines, fill=dotorange]
      (\x,\y) circle (\bulletsize);
}
\end{tikzpicture}
\;.
\label{eq:k_stack_diagram}
\end{equation}
On the left, the brace marks the $k$ stacked replicas (alternating
$\mathbf{U}$ in sky-blue and $\mathbf{U}^*$ in light-steel-blue,
following~Eqs.~\eqref{eq:rmpsBrickwall}--\eqref{eq:rmpsBrickwall_conj}),
each running for $t$ brickwork layers in the time direction.  On the
right, the $k$ replicas have collapsed into a \emph{single} 2D
tensor network: the bricks denote the Haar-averaged gates
$\widetilde{\mathcal{T}}^{(k)}_{i, i+1}$, while orange dots on every
inter-layer wire are the $\mathrm{S}_k$-spin contractions. 
We now detail how these elements arise.

\subsubsection{Averaging over Haar-random gates}
\label{subsec:weingarten}

By statistical independence, the key problem is to compute $k$th moment of a gate drawn with Haar measure on $\mathrm{U}(q)$. This is given by Schur-Weyl duality~\cite{mele_introduction_2024}
\begin{equation}
\mathbb{E}_{U \sim \mathrm{Haar}}\bigl[(U \otimes U^*)^{\otimes k}\bigr]
\;=\;
\sum_{\sigma, \pi \in \mathrm{S}_k}\,\mathrm{Wg}_{\sigma, \pi}(q)\,|\sigma\rrangle\llangle\pi| ,
\label{eq:haar_moment}
\end{equation}
expressed in the replica-spin basis~Eq.~\eqref{eq:perm_state}.
The coefficient $\mathrm{Wg}_{\sigma, \pi}(q)$ is the
\emph{Weingarten function}, defined as the (Moore--Penrose)
pseudo-inverse of the Gram matrix $G(q)$ with elements $G_{\sigma, \pi}(q)=\llangle \sigma|\pi\rrangle$, cf.~Eq.~\eqref{eq:gram_def}.
The pseudo-inverse is required only at
$q < k$, where the permutation states cease to be linearly
independent. We will work only the regime $q\ge k$ throughout these notes, so the inverse can be taken swiftly.

In the following, both the Gram and Weingarten matrices will appear several times for $k=2$. For this reason, we present in detail their expressions. Recall that the symmetric
group for $k=2$ is $\mathrm{S}_2 = \{\iota, s\}$, with $\iota$ the identity and
$s = (1\,2)$ the transposition.  
By direct inspection using Eq.~\eqref{eq:iota_swap_states}, we contruct $G_{\sigma, \pi}(q)=\llangle \sigma|\pi\rrangle = q^{\#(\sigma^{-1}\pi)}$, where $\#(\sigma)$ is the number of cycles in the permutation $\sigma$.
In cycle notation $\iota =()\equiv  (1)(2)$ has $\#(\iota) = 2$ cycles, while $s = (1\,2)$ has $\#(s) = 1$ cycle,
so the Gram matrix takes only two values: $q^2$ on the diagonal
($\sigma = \pi$, $\sigma^{-1}\pi = \iota$) and $q$ on the
off-diagonal ($\sigma \neq \pi$, $\sigma^{-1}\pi = s$).  

Hence
\begin{equation}
G(q) = \begin{pmatrix} \llangle \iota|\iota\rrangle & \llangle \iota|s\rrangle \\ \llangle s|\iota\rrangle & \llangle s|s\rrangle\end{pmatrix}
= \begin{pmatrix} q^2 & q \\ q & q^2 \end{pmatrix} ,
\label{eq:gram_Sk_k2}
\end{equation}
whose determinant is $\det G(q) = q^4 - q^2 = q^2(q^2 - 1)$,
non-vanishing for any $q \geq 2$.  Inverting,
\begin{equation}
\mathrm{Wg}(q) = G(q)^{-1}
= \frac{1}{q^2(q^2 - 1)}\begin{pmatrix} q^2 & -q \\ -q & q^2 \end{pmatrix} ,
\label{eq:wg_k2}
\end{equation}
and one verifies $G(q)\,\mathrm{Wg}(q) = \mathbb{I}_{2\times 2}$ by
direct multiplication.  
For the random gate of interest $q=d^2$, and Eq.~\eqref{eq:wg_k2} gives the exact closed form
$\mathrm{Wg}(d^2) = \tfrac{1}{d^2(d^4-1)}\bigl(\begin{smallmatrix} d^2 & -1 \\ -1 & d^2 \end{smallmatrix}\bigr)$. 

\begin{exercise}
Assuming $q>3$, compute the Gram matrix $G(q)$ and the Weingarten matrix
$\mathrm{Wg}(q) = G^{-1}(q)$ for $k = 3$, by repeating the construction
of~Eqs.~\eqref{eq:gram_Sk_k2}--\eqref{eq:wg_k2}.
\textbf{(a)} List the six elements of $\mathrm{S}_3$ and their cycle
counts.  
\textbf{(b)} Cross-check your answer numerically with
\verb|gram_matrix(SymmetricBasis(3), q)| and
\verb|weingarten_matrix(SymmetricBasis(3), q)|.
\end{exercise}

\subsubsection{Global Haar random unitaries and states}
\label{subsec:weingarten}
The simplest non-trivial application of~Eq.~\eqref{eq:haar_moment} is
a \emph{single} Haar-random unitary $U \in \mathrm{U}(D)$ on the
full Hilbert space, with $D = d^N$ and the system starting from the
product reference state
\begin{equation}
|\rho_0^{\otimes k}\rrangle \;=\; (\mathbb{I}\otimes |\pmb{0}\rangle\langle \pmb{0}|^{\otimes k})|\varphi\rangle \;\equiv\; |0^{\otimes 2kN}\rrangle ,
\label{eq:zero_kk}
\end{equation}
where the right-most form is the shorthand we use throughout for the
all-zero ket on the $k$-replicated doubled space. Note also that $\mathfrak{A}|0^{\otimes 2kN}\rrangle=|0^{\otimes 2kN}\rrangle$ under reshaping.
Substituting the Haar moment~Eq.~\eqref{eq:haar_moment} into the IPR template~Eq.~\eqref{eq:ipr_replica} yields
\begin{equation}
\begin{split}
\mathbb{E}_\mathrm{Haar}\!\bigl[\mathcal{I}^{(k)}\bigr] 
\;&=\;
\sum_{\sigma, \pi \in \mathrm{S}_k} \mathrm{Wg}_{\sigma, \pi}(D)\,
   \llangle \Omega^{\textup{IPR}}_k | \sigma\rrangle\,\llangle \pi | 0^{\otimes 2kN}\rrangle  \\
   & =\sum_{\sigma, \pi \in \mathrm{S}_k} \mathrm{Wg}_{\sigma, \pi}(D)\,
   \prod_{i,j=1}^N\llangle \Omega^{\textup{IPR}}_k | \sigma\rrangle_i\,\llangle \pi | 0^{\otimes 2k}\rrangle_j,
\label{eq:ipr_haar_general}
\end{split}
\end{equation}
as in the reshaped base the permutation state is a product state $|\sigma\rrangle=\bigotimes_{i=1}^N |\sigma\rrangle_i$ acting on each individual qudit.

The two boundary overlaps simplify dramatically.  Recall that the IPR boundary factorises over sites. Thus, we need to compute
\begin{equation}
\llangle \Omega^{\textup{IPR}}_k | \sigma\rrangle_i \;=\;\sum_{x}\langle x,x,x,x|\sigma\rrangle= d,
\qquad
\sigma \in \mathrm{S}_k ,
\label{eq:ipr_overlap_site}
\end{equation}
since $\langle x,x,x,x|$ kills the sum in $|\iota\rrangle$  and $|s\rrangle$. 
This implies that for the full chain we have 
\begin{equation}
\llangle \Omega^{\textup{IPR}}_k | \sigma\rrangle  = \prod_{i=1}^N \llangle \Omega^{\textup{IPR}}_k | \sigma\rrangle_i\;=\; d^N \;=\; D
\label{eq:ipr_overlap_global}
\end{equation}
for every $\sigma \in \mathrm{S}_k$, independent of $\sigma$.  
Similarly, since the state $|0^{\otimes 2k}\rrangle$ is permutational invariant, we have 
\begin{equation}
\llangle \pi | 0^{\otimes 2kN}\rrangle \;=\prod_{j=1}^N \llangle \pi | 0^{\otimes 2k}\rrangle_j\; =1 ,
\qquad
\pi \in \mathrm{S}_k .
\label{eq:init_overlap}
\end{equation}
The double sum in~Eq.~\eqref{eq:ipr_haar_general} therefore collapses
to the total Weingarten row-sum, multiplied by $D$:
\begin{equation}
\mathbb{E}_\mathrm{Haar}\!\bigl[\mathcal{I}^{(k)}\bigr]
\;=\;
D\,\sum_{\sigma, \pi \in \mathrm{S}_k}\mathrm{Wg}_{\sigma, \pi}(D)
\;=\;
\frac{k!\,D!}{(D + k - 1)!} ,
\qquad D = d^N ,
\label{eq:ipr_haar}
\end{equation}
where the closed form on the right comes from the standard
identity~\cite{collins_weingarten_2022} 
\begin{equation}
\sum_{\sigma, \pi \in \mathrm{S}_k}\mathrm{Wg}_{\sigma, \pi}(D)
   \;=\;
   \frac{k!\,(D - 1)!}{(D + k - 1)!} .
\end{equation}

As a pedagogical check, let us verify
Eq.~\eqref{eq:ipr_haar} from scratch at $k = 2$.  The Weingarten
matrix is read from Eq.~\eqref{eq:wg_k2} at $q = D$:
\begin{equation}
\mathrm{Wg}(D)
\;=\;
\frac{1}{D^2 (D^2 - 1)}
\begin{pmatrix} D^2 & -D \\ -D & D^2 \end{pmatrix} ,
\label{eq:wg_global_k2}
\end{equation}
so the four entries of the row-sum
$\sum_{\sigma, \pi \in \mathrm{S}_2}\mathrm{Wg}_{\sigma, \pi}(D)$ split into
a diagonal and an off-diagonal contribution,
\begin{equation}
\begin{aligned}
\mathrm{Wg}_{\iota \iota} + \mathrm{Wg}_{s s}
   &= \frac{2 D^2}{D^2 (D^2 - 1)} = \frac{2}{D^2 - 1} , \\
\mathrm{Wg}_{\iota s} + \mathrm{Wg}_{s \iota}
   &= -\frac{2 D}{D^2 (D^2 - 1)} = -\frac{2}{D(D^2 - 1)} .
\end{aligned}
\end{equation}
Adding the two we get,
\begin{equation}
\sum_{\sigma, \pi \in \mathrm{S}_2}\mathrm{Wg}_{\sigma, \pi}(D)
\;=\;
\frac{2}{D^2 - 1} \,-\, \frac{2}{D(D^2 - 1)}
\;=\;
\frac{2(D - 1)}{D(D^2 - 1)}
\;=\;
\frac{2}{D(D + 1)} ,
\label{eq:wg_rowsum_k2}
\end{equation}
where we used $D^2 - 1 = (D - 1)(D + 1)$, as we wished to prove.  
Multiplying by~$D$ from~Eq.~\eqref{eq:ipr_haar} gives the textbook ``Page IPR''
result
\begin{equation}
\mathcal{I}_\mathrm{Haar}^{(2)}:=\mathbb{E}_\mathrm{Haar}\!\bigl[\mathcal{I}^{(2)}\bigr]
\;=\;
\frac{2}{D + 1}
\;\xrightarrow{D\,\to\,\infty}\;
\frac{2}{D} \;.
\label{eq:ipr_haar_k2}
\end{equation}
As anticipated, Haar-random states have collision probability scaling as $2/D$. The extra factor of two is a genuinely quantum-correlation effect: if the bits were drawn independently, one would instead obtain a classical mixture with collision probability $1/D$, corresponding to the uniform distribution over $D=d^N$ outcomes.

At $k=2$, the very fact that an ensemble of states has a typical collision probability that matches the 
IPR of random states $\mathcal{I}_\mathrm{Haar}^{(2)}$ is the definition of \emph{anticoncentration}~\cite{dalzell2022random}.  
More generally, when the $k$th IPR of a circuit ensemble
attains its Haar value~Eq.~\eqref{eq:ipr_haar},
$\mathbb{E}\bigl[\mathcal{I}^{(k)}\bigr]
   \approx \mathcal{I}_\mathrm{Haar}^{(k)}$,
we say the ensemble is \emph{anticoncentrated at level $k$}: the
agreement extends to the entire $k$th moment of the output
distribution, and increasing $k$ progressively constrains higher
fluctuation modes of the
distribution~\cite{lami2024anticoncentration,magni2025anticoncentration,christopoulos2024universal}.
The $k = 2$ condition underpins the complexity-theoretic arguments
for quantum computational
advantage~\cite{boixo2018characterizing,arute2019quantum,jens}:
Once the output $|\Psi\rangle$ of a noiseless brickwork has been anticoncentrated, no
classical algorithm is believed to \textit{sample} from $p(\pmb{x})=|\langle \pmb{x}|\Psi\rangle|^2$ in any polynomial time.  
How fast a finite-depth brickwork relaxes to its anticoncentrated plateau, the so-called anticoncentration timescale $t_{\rm AC}(N)$, is one of the core quantitative outputs of the replica tensor-network construction we develop in the following sections.

The same one-gate calculation gives the $k$th local purity
$\mathcal{P}^{(k)}_A$ on a region $A$ with $|A| = \ell$.  The bulk
Haar moment is identical; only the top boundary changes,
$\Omega^{\textup{IPR}}_k \to \Omega^{\textup{LP}}_k$ of
Eq.~\eqref{eq:boundary_LP}, and the per-site overlap
$\llangle \Omega^{\textup{LP}}_k | \sigma\rrangle$ now depends on
whether the site lies in $A$ (factor $d^{\#(\eta\sigma)}$, with
$\eta$ the cyclic permutation) or in $B$ (factor $d^{\#(\sigma)}$).
At $k = 2$, a direct $2 \times 2$ contraction yields the celebrated
Page value~\cite{page,Zyczkowski2001}
\begin{equation}
\mathbb{E}_\mathrm{Haar}\!\bigl[\mathcal{P}^{(2)}_A\bigr]
\;=\;
\frac{D_A + D_B}{D_A D_B + 1} ,
\qquad D_A = d^{\ell},\ D_B = d^{N - \ell} ,
\label{eq:page_value}
\end{equation}
which behavers for $D_A \ll D_B$ as $\mathcal{P}^{(2)}_A \approx 1/D_A$ (the partial trace
$\rho_A$ is approximately maximally mixed), while for $D_A \simeq D_B$ 
it gives $\mathcal{P}^{(2)}_A \approx 2 / D_A$ (the small correction accounts are the Page corrections). 

\begin{exercise}
Complete the derivation of Eq.~\eqref{eq:page_value}, and generalize the result for  $\mathbb{E}_\mathrm{Haar}\bigl[\mathcal{P}^{(3)}_A\bigr]$ for
a Haar random state.  \textbf{(a)} Express the per-site
overlap of the LP boundary with $\sigma \in \mathrm{S}_3$ in terms of the
cycle counts of $\sigma$ and of $\eta\sigma$, where
$\eta = (1\,2\,3)$ is the cyclic permutation.  \textbf{(b)} Reduce
the closed form to a sum over $\mathrm{S}_3$ of Weingarten coefficients with
the boundary weights, and study the leading large-$D$ behaviour at
fixed $\ell / N$.  
\end{exercise}
\subsubsection{The replica transfer matrix}
\label{subsec:transfer}

Applying the Haar average~Eq.~\eqref{eq:haar_moment} to each gate $U_{i,i+1}$ in the circuit, we obtain a \emph{replica transfer matrix} $\mathcal{T}^{(k)}_{i,i+1}$:
\begin{equation}
\mathcal{T}^{(k)}_{i,i+1} = \begin{tikzpicture}[baseline=(current bounding box.center), scale=0.85]
  \definecolor{wgBlue}{RGB}{52,101,164}
  \definecolor{darkgraylines}{RGB}{55,55,55}
  \definecolor{dotorange}{RGB}{245,130,32}
  \definecolor{crimsonred}{RGB}{66,135,245}    
\definecolor{darkgraylines}{RGB}{55,55,55}
\definecolor{dotorange}{RGB}{245,130,32}
  \draw[thick, darkgraylines] (-0.4,-0.2) -- (-0.4, 0.75);
  \draw[thick, darkgraylines] ( 0.4,-0.2) -- ( 0.4, 0.75);
  \draw[ultra thick, fill=crimsonred, rounded corners=2pt]
     (-0.6, 0.05) rectangle (0.6, 0.55);
\end{tikzpicture} =\sum_{\sigma, \pi \in \mathrm{S}_k} \mathrm{Wg}_{\sigma,\pi}(d^2)\, |\sigma\rrangle_i |\sigma\rrangle_{i+1} \llangle \pi|_i \llangle \pi|_{i+1}\;,
\label{eq:transfer_basic}
\end{equation}
where in the second step we use the factorization property of permutation states on the replica space of individual qudits. 
Within the permutation states, the transfer matrix acts as a $(k!)\times (k!)$ matrix. 
In fact, Eq.~\eqref{eq:transfer_basic} is a projection onto the space spanned by permutation operators $\{|\sigma\rrangle_i|\sigma\rrangle_{i+1}\}$. 

To build the full tensor network, we must account for the fact that adjacent layers share sites and that permutation operations are not orthogonal. 
Concretely, between two consecutive layers, the permutation states on the shared site are contracted via the Gram matrix:
\begin{equation}
G_{\sigma,\tau}(d) = \llangle \sigma|\tau\rrangle = d^{\#(\sigma^{-1}\tau)}.
\label{eq:gram}
\end{equation}

In principle, one now has all the ingredients in Eq.~\eqref{eq:k_stack_diagram} and could implement the brickwork RTN directly using the raw transfer matrix $\mathcal{T}^{(k)}_{i, i+1}$ of
Eq.~\eqref{eq:transfer_basic} and applying the Gram contractions between consecutive layers as separate steps.  
In practice, this is numerically unstable: every layer multiplies the spin amplitudes by factors of $\sim d^{N}$, which grow exponentially in $N$ while the Weingarten coefficients carry compensating exponentially small factors $\sim d^{-N}$.  Storing them separately accumulates catastrophic cancellation, so the running MPS bond amplitudes become either astronomically large or numerically zero after a handful of layers.

To resolve this problem, we absorb these overlaps into the definition of the tensor by introducing the ``dressed'' transfer matrix
\begin{equation}
\begin{split}
\widetilde{\mathcal{T}}^{(k)}_{i,i+1}
\;&\equiv\;
\begin{tikzpicture}[baseline=(current bounding box.center), scale=0.85]
  \definecolor{wgBlue}{RGB}{52,101,164}
  \draw[thick] (-0.4,-0.75) -- (-0.4,-0.25);
  \draw[thick] ( 0.4,-0.75) -- ( 0.4,-0.25);
  \draw[thick] (-0.4, 0.75) -- (-0.4, 0.25);
  \draw[thick] ( 0.4, 0.75) -- ( 0.4, 0.25);
  \draw[ultra thick, fill=wgBlue, rounded corners=2pt]
     (-0.6,-0.25) rectangle (0.6, 0.25);
\end{tikzpicture}
\;=\;
\begin{tikzpicture}[baseline=(current bounding box.center), scale=0.85]
  \definecolor{wgBlue}{RGB}{52,101,164}
  \definecolor{darkgraylines}{RGB}{55,55,55}
  \definecolor{dotorange}{RGB}{245,130,32}
  \definecolor{crimsonred}{RGB}{66,135,245}    
\definecolor{darkgraylines}{RGB}{55,55,55}
\definecolor{dotorange}{RGB}{245,130,32}
  \draw[thick, darkgraylines] (-0.4,-1.05) -- (-0.4, 0.75);
  \draw[thick, darkgraylines] ( 0.4,-1.05) -- ( 0.4, 0.75);
  \draw[ultra thick, fill=crimsonred, rounded corners=2pt]
     (-0.6, 0.05) rectangle (0.6, 0.55);
  \draw[thick, draw=darkgraylines, fill=dotorange]
     (-0.4,-0.45) circle (0.10);
  \draw[thick, draw=darkgraylines, fill=dotorange]
     ( 0.4,-0.45) circle (0.10);
\end{tikzpicture}
\\[0.4em]
\;&\equiv\;
\sum_{\substack{\sigma,\pi \in \mathrm{S}_k \\ \pi_1,\pi_2 \in \mathrm{S}_k}}
   \mathrm{Wg}_{\sigma,\pi}(d^2)\,
   G_{\pi,\pi_1}(d)\, G_{\pi,\pi_2}(d)\,
   |\sigma\rrangle_i |\sigma\rrangle_{i+1}\llangle \check\pi_1|_i  \llangle\check\pi_2|_{i+1}
   ,
\end{split}
\label{eq:transfer_dressed}
\end{equation}
where the input legs carry the dual basis $| \check\pi_j\rrangle$ of the symmetric-group commutant, defined by $\llangle \check\pi|\tau\rrangle = \llangle \tau|\check\pi\rrangle=\delta_{\pi,\tau}$, and the on-site Gram factors $G_{\pi,\pi_j}(d)$ contract the layer-output $\sigma$ at site $i$ ($i+1$) with the layer-input $\pi_j$ that the next layer above expects. 
This dressed transfer matrix folds the $\mathrm{Wg}$ and $G$ factors into a \emph{single} four-index tensor whose entries are of order one (Weingarten and Gram cancel each other inside this tensor), making the contraction feasible at $N \gg 1$. 
We will employ the symbol $\widetilde{\mathcal{T}}^{(k)}$ throughout the rest of these notes.  

At $k = 2$, the dressed transfer
matrix~Eq.~\eqref{eq:transfer_dressed} can be written down explicitly
on the basis $\{|\iota \iota\rrangle, |\iota s\rrangle, |s \iota\rrangle, |s s\rrangle\}$
of two-site replica spins.  The non-zero matrix elements come from the
two ferromagnetic configurations $|\iota \iota\rrangle$ and $|s s\rrangle$,
which propagate to themselves with weight $1$, plus the two ``mixed''
states $|\iota s\rrangle$ and $|s \iota\rrangle$, which feed back into
$|\iota \iota\rrangle$ and $|s s\rrangle$ with the domain-wall weight
$K_d = d/(d^2 + 1)$:
\begin{equation}
\widetilde{\mathcal{T}}^{(2)}_{i, i+1}
\;=\;
\begin{pmatrix}
   1 & K_d & K_d & 0 \\
   0 & 0   & 0   & 0 \\
   0 & 0   & 0   & 0 \\
   0 & K_d &K_d & 1
\end{pmatrix} ,
\qquad
K_d \;=\; \frac{d}{d^2 + 1} .
\label{eq:transfer_T22_explicit}
\end{equation}
This is the building block we use throughout the IPR and local-purity
worked examples of \S\ref{sec:applications_1}--\ref{sec:applications_1_LP},
and the same matrix governs the random-walk solution of the
half-chain purity in Eq.~\eqref{eq:P2_exact} below.
It is worth noting that the local Hilbert space dimension, daunting direct implementations in vector simulations or standard tensor network evolution, enters here as a parameter. This is a nice feature of replica tensor networks that recurs in several settings of interest.

\subsubsection{Boundary conditions}
\label{sec:rtn:boundaries}
Once obtained the average bulk transfer matrix in Eq.~\eqref{eq:k_stack_diagram}, we need to impose the boundary conditions that come from the initial state and the observable of interest.

For the product initial state $|\Psi_0\rangle = |0\rangle^{\otimes N}$, the per-site overlap between the reference vector $|\mathfrak{I}^{(k)}\rrangle=\mathfrak{A}|\rho_0^{\otimes k}\rrangle = |0^{\otimes 2kN}\rrangle$ and any permutation ket is $\llangle \sigma | 0^{\otimes 2k}\rrangle_i = 1$, independent of $\sigma$.  Contracting the first layer of gates in Eq.~\eqref{eq:k_stack_diagram} with the initial state results into the lower replica boundary condition $|\mathfrak{I}^{(k)}\rrangle=\bigotimes_{i=1}^{N/2}|\mathfrak{I}_{2i-1,2i}\rrangle$.
Each of these states is obtained by acting with an individual replica gate $\mathcal{T}_{i,i+1}$ and using the Weingarten sum rule. 
A simple computation leads to 
\begin{equation}
\begin{split}
\kket{\mathfrak{I}_{i,i+1}}
\;&\equiv\;
\begin{tikzpicture}[baseline=(current bounding box.center), scale=0.85]
\definecolor{initYel}{RGB}{240,188,66}
\draw[thick] (-0.4, 0.25) -- (-0.4, 0.75);
\draw[thick] ( 0.4, 0.25) -- ( 0.4, 0.75);
\draw[ultra thick, fill=initYel, rounded corners=2pt]
   (-0.6,-0.25) rectangle (0.6, 0.25);
\node at (0, 0) {\small $+$};
\end{tikzpicture}
\;=\;
\mathcal{T}^{(k)}_{i,i+1}\, |\rho_0^{\otimes k}\rrangle
\;=\;
\frac{1}{\prod_{m=0}^{k-1}(d^2+m)}
\sum_{\sigma \in \mathrm{S}_k}\, |\sigma\rrangle_i |\sigma\rrangle_{i+1} ,
\end{split}
\label{eq:bc_init}
\end{equation}
in which all permutations enter with equal weight, reflecting the
maximal symmetry of $|\Psi_0\rangle$ under replica permutations. The
prefactor is the one-gate Haar integral on $\mathrm{U}(d^2)$~\cite{collins_weingarten_2022}.
This state captures the first time step of the replica evolution $t=1$.

At the top of the strip, we introduce the \emph{observable dependent} replica boundary $|\Omega_k^\Lambda\rrangle$.
Both the IPR and the local purity factorize, thus $|\Omega_k^\Lambda\rrangle=\bigotimes_{i=1}^N |\Omega_k^\Lambda\rrangle_i$ for the appropriately chosen replica qudit states
\begin{equation}
\begin{split}
\bbra{\Omega_k^\Lambda}_i
\;&\equiv\;
\begin{tikzpicture}[baseline=(current bounding box.center), scale=0.85]
  \definecolor{topOra}{RGB}{255,102,0}
  \draw[thick] (-0.3,-0.75) -- (-0.3,-0.25);
  \draw[ultra thick, fill=topOra, rounded corners=2pt]
     (-0.6,-0.25) rectangle (0.0, 0.25);
  \node at (-0.3, 0) {\small $\hat{\Lambda}$};
\end{tikzpicture}
\;\equiv\;
\sum_{\sigma\in \mathrm{S}_k} b_\Lambda(\sigma)\, \bbra{\check\sigma} .\end{split}
\label{eq:noisy_top_box}
\end{equation}
In the above expression, $b_\Lambda(\sigma) = \llangle \Omega_k^\Lambda | \sigma\rrangle_i$ is
the per-site observable amplitude of \S\ref{sec:rtn:boundaries}. 
The orange box thus carries \emph{all} the
observable-specific data through the
weights $b_\Lambda(\sigma)$.

Let us discuss concretely the examples for the inverse participation ratios ($\Lambda=\mathrm{IPR}$) and the local purity ($\Lambda=\mathrm{LP}$). 
The IPR boundary
$|\Omega^{\textup{IPR}}_k\rrangle$ has already been written
in~Eq.~\eqref{eq:boundary_IPR}.  
Its per-site overlap with any
$\sigma$-spin is
\begin{equation}
\llangle \Omega^{\textup{IPR}}_k | \sigma\rrangle_i \;=\; d ,
\qquad
\sigma \in \mathrm{S}_k ,
\label{eq:bc_ipr_flat}
\end{equation}
because the diagonal projector forces all $k$ ket-bra pairs on the
site to take the same computational basis value $x_i$, and summing
over $x_i \in \{0, \ldots, d - 1\}$ produces a single factor of $d$
that is \emph{independent of $\sigma$}.

For the LP boundary
$|\Omega^{\textup{LP}}_k\rrangle = \bigotimes_{i\in A}|\iota\rrangle_i \otimes \bigotimes_{i\in B}|\eta\rrangle_i$
of~Eq.~\eqref{eq:boundary_LP_kets}, the per-site overlap with a
$\sigma$-spin reads
\begin{equation}
\llangle \Omega^{\textup{LP}}_k | \sigma\rrangle_i
\;=\;
\begin{cases}
   G_{\iota, \sigma}(d) = d^{\#(\sigma)} & i \in B , \\[2pt]
   G_{\eta, \sigma}(d) = d^{\#(\eta^{-1}\sigma)} & i \in A ,
\end{cases}
\label{eq:bc_LP_flat}
\end{equation}
i.e. it is exactly the single-site Gram matrix
entry, cf. Eq.~\eqref{eq:gram_def}, between the $\sigma$ at site $i$ and the
fixed boundary permutation ($\eta$ on $A$, $\iota$ on $B$).

We conclude with a note on implementation and good practice. To avoid overflow or underflow caused by constant prefactors in the initial and final boundary conditions, it is convenient to factor them out explicitly in the code. 
For the initial state, we remove the factor $\prod_{m=0}^{k-1}(d^2+m)$ by using the rescaled bottom boundary condition $|\tilde{\mathfrak{I}}_{i,i+1}\rrangle = \sum_{\sigma\in \mathrm{S}_k} |\sigma\rrangle_i|\sigma\rrangle_{i+1}$. Similarly, for the observable, we factor out $d$ for the IPR and $d^2$ for the LP, and implicitly define $|\tilde{\Omega}_k^\Lambda\rrangle = \sum_{\sigma\in \mathrm{S}_k} \tilde b_\Lambda(\sigma)|\check\sigma\rrangle$. Keeping track of these constants separately can be crucial for reaching larger system sizes. Thus, in the numerical implementation of Sec.~\ref{sec:rtn:mps_evolution}, we compute the bulk contraction with \emph{unit} per-site weights and multiply by the appropriate overall constants only when extracting the final result.

\paragraph{Putting it all together.}~The full replica tensor network for the $k$th averaged observable is obtained by stacking the dressed bulk tensors $\widetilde{\mathcal{T}}^{(k)}$ of Eq.~\eqref{eq:transfer_dressed} (blue boxes) between the flat initial boundary of Eq.~\eqref{eq:bc_init} (yellow boxes) and the observable-dependent top boundary of Eq.~\eqref{eq:noisy_top_box} (orange boxes):
\begin{equation}
  \mathbb{E}\bigl[\Lambda(|\Psi_t\rangle)\bigr]
\;=\;
\llangle \Omega_k^{\Lambda}\,\big|\,\Phi_t^{(k)}\,
   \big|\mathfrak{I}^{(k)}\rrangle
  \;=\;
  \begin{tikzpicture}[baseline=(current bounding box.center), scale=0.42]
    \definecolor{wgBlue}{RGB}{52,101,164}
    \definecolor{initYel}{RGB}{240,188,66}
    \definecolor{topOra}{RGB}{255,102,0}
    \draw[decorate,decoration={brace},thick] (-0.6,-4.5) -- node[left]{$t$} (-0.6,5.7);
    \foreach \i in {1,...,6}{ \draw[thick] (\i,-4.5)--(\i,5.5); }
    \foreach \i in {1,3,5}{
      \draw[ultra thick, fill=initYel, rounded corners=2pt]
         (\i-0.4,-4.5) rectangle (\i+1.4,-3.7);
      \node at (\i+0.5,-4.10) {\tiny $+$};
    }
    \foreach \jj/\stops in {0/{1,3,5},1/{2,4},2/{1,3,5},3/{2,4},4/{1,3,5},5/{2,4}}{
      \pgfmathsetmacro{\yy}{-2.5 + 1.20*\jj}
      \foreach \i in \stops{
        \draw[ultra thick, fill=wgBlue, rounded corners=2pt]
           (\i-0.4,\yy-0.40) rectangle (\i+1.4,\yy+0.40);
      }
    }
    \foreach \i in {1,...,6}{
      \draw[ultra thick, fill=topOra, rounded corners=2pt]
         (\i-0.4,4.7) rectangle (\i+0.4,5.5);
      \node at (\i,5.10) {\tiny $\hat{\Lambda}$};
    }
  \end{tikzpicture}\,.
  \label{eq:rtn_full}
\end{equation}
The strip is read bottom-to-top: the yellow row prepares the flat initial kets $\kket{\mathfrak{I}_{i,i+1}}$, the blue brickwork layers implement the dressed two-site averaged gates $\widetilde{\mathcal{T}}^{(k)}$, and the orange row applies the final flat bras $\bbra{\Omega_k^\Lambda}_i$ encoding the observable $\Lambda$. The brace on the left counts the circuit depth $t$. Diagrams of this form are the central object of the rest of these notes; Sec.~\ref{sec:rtn:mps_evolution} then gives the algorithmic recipe for their contraction.

\subsection{Replica tensor network as MPS evolution}
\label{sec:rtn:mps_evolution}

\label{sec:rtn}
The whole object in~Eq.~\eqref{eq:rtn_full} is a classical statistical-mechanics model with $k!$-state spins on a two-dimensional lattice, in which the (in general non-positive) ``Boltzmann weights'' are the Weingarten functions $\mathrm{Wg}_{\sigma, \pi}(d^2)$ and the Gram entries $G_{\sigma, \pi}(d) = d^{\#(\sigma^{-1}\pi)}$.
This observation is at the heart of key results in condensed matter theory, e.g. on entanglement spreading in Ref.~\cite{zhou2}, and in quantum information such as anticoncentration of random circuits, cf. Ref.~\cite{dalzell2022random}. 
We now explain how to contract Eq.~\eqref{eq:rtn_full} as a tensor network on a two-dimensional $N\times t$ lattice. The key observation is that this network can be read as a non-unitary evolution in the \emph{space} direction.

\subsubsection{Contraction as MPS evolution}

At each fixed depth, the row of $\sigma$-spins
is encoded as a matrix product state (MPS) on a chain of $k!$-state
``super-sites'',
\begin{equation}
|\Pi_t\rrangle \;=\;
   \sum_{\sigma_1, \ldots, \sigma_N \in \mathrm{S}_k}
     A_{[1]}^{\sigma_1} A_{[2]}^{\sigma_2} \cdots A_{[N]}^{\sigma_N}\,
     |\sigma_1, \ldots, \sigma_N\rrangle ,
\label{eq:mps}
\end{equation}
with $\chi \times \chi$ tensors $A^{\sigma}_{[i]}$ (with the open
boundary tensors having $\chi = 1$ on the outer leg) and bond
dimension $\chi$ that we are about to bound.  At time $t = 0$, this is the initial state $|0^{2kn}\rrangle$, whereas at $t=1$ it corresponds to the replica boundary in~Eq.~\eqref{eq:bc_init}.

A single brickwork layer of the network corresponds to the application
of a matrix product operator (MPO) on this MPS, built by stacking the
even-bond and odd-bond rows of dressed transfer matrices
$\widetilde{\mathcal{T}}^{(k)}_{i, i+1}$.
This is just the time-evolved block decimation~(TEBD) algorithm~\cite{Schollwoeck11,pollo}, which we sketch here and briefly review in Appendix~\ref{sec:implementation_observables}. 
In a nutshell, there are four ingredients to consider: (i) canonical-form sweep, (ii) two-site gate contraction, (iii) SVD truncation, (iv) boundary contraction. 
We carry out the layer by sweeping over bonds left to right: at each bond $(i, i+1)$ we move the orthogonality centre of the MPS to site $i$ via successive QR decompositions, contract the four-leg gate $\widetilde{\mathcal{T}}^{(k)}_{i, i+1}$ with the two on-site tensors $A_{[i]}, A_{[i+1]}$ into a single rank-four block,  reshape and SVD-truncate that block back to two MPS tensors with bond dimension $\chi_{\rm new} \leq \chi_{\max}$, and  re-absorb the singular values into the right tensor so that the orthogonality centre advances by one site.  
For a fixed $\chi$, the cost of one layer is therefore $\Lambda(N \cdot (k!)^3 \cdot \chi^3)$. 
Crucially, because of non-unitarity, $\chi_{\max} = O(\mathrm{poly}(k!))$ is often sufficient for convergence, irrespective of the system size. 
After $t$ such layers, the average observable is the inner product of the evolved MPS with the top boundary $|\Omega^{\Lambda}_k\rrangle$ of \S\ref{subsec:transfer}.  

\subsubsection{Worked example: anticoncentration of Haar brickworks}
\label{sec:applications_1}
\label{sec:hands_on_1}

In this first hands-on subsection, we turn the tensor-network formalism of Sec.~\ref{sec:rtn:mps_evolution} into a concrete numerical pipeline. We focus on the inverse participation ratio of a Haar-unitary brickwork circuit on $N$ qubits, starting from $k=2$, i.e.~the collision probability, and returning later to the case $k=3$.

The companion library \texttt{ReplicaTN} presents a direct implementation of what is presented in these notes. It uses two main ingredients: a commutant basis \texttt{B} and a boundary specification \texttt{bd}. For unitary Haar dynamics at $k=2$ on qubits, the minimal setup is
\begin{pybox}
import replicatn as rtn

B   = rtn.SymmetricBasis(2)       # |B| = 2!  (unitary Haar, k = 2)
d   = 2                           # qubits
bd  = rtn.IPRBoundary(B, d)       # b_i(sigma) = d for every sigma
val = rtn.brickwork_average(B, d, N, t, bd)
\end{pybox}

\begin{funcbox}{SymmetricBasis(k)}
Enumerates the $k!$ permutations in $\mathrm{S}_k$, represented in one-line notation, and stores the corresponding commutant basis. This basis is then reused in the construction of the Gram matrix, the Weingarten matrix, and the averaged local tensors. Throughout this subsection we use $k=2$; switching to $k=3$ only requires replacing \texttt{SymmetricBasis(2)} by \texttt{SymmetricBasis(3)}.
\end{funcbox}

\begin{figure}[t!]
\centering
\includegraphics[width=\textwidth]{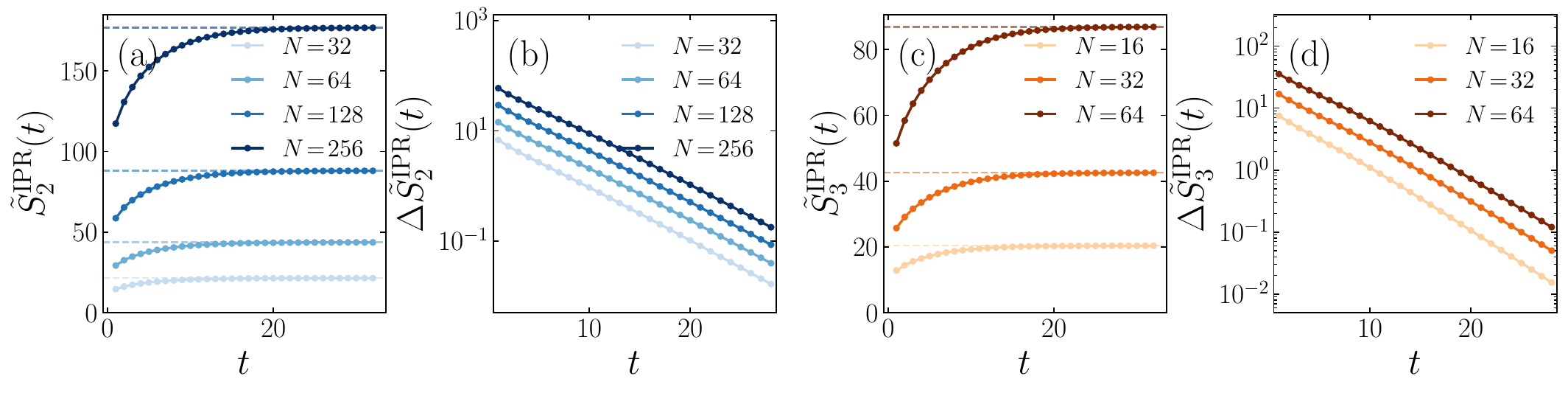}
\caption{Annealed inverse participation ratios for Haar-unitary qubit brickwork circuits. \emph{(a)} Annealed participation entropy $\widetilde{S}^{\,\rm IPR}_{2}(t) \equiv -\log \mathbb{E}[\mathcal{I}^{(2)}]$ as a function of depth $t$ for $N \in \{32,64,128,256\}$ (Blues palette). Dashed horizontal lines mark the Haar plateau $\widetilde{S}^{\,\rm IPR}_{2,\rm Haar} = -\log \mathcal{I}^{(2)}_{\mathrm{Haar}}$ of Eq.~\eqref{eq:ipr_haar_k2}. \emph{(b)} Relative deviation from the Haar plateau, $(\mathbb{E}[\mathcal{I}^{(2)}]-\mathcal{I}^{(2)}_{\mathrm{Haar}})/\mathcal{I}^{(2)}_{\mathrm{Haar}}$, shown on a logarithmic scale. The decay is exponential, $\propto N e^{-\alpha_2 t}$. \emph{(c,d)} Same analysis for $k=3$ and $N \in \{16,32,64\}$ (Oranges palette), with Haar value $\mathcal{I}^{(3)}_{\mathrm{Haar}} = 6/[(D+1)(D+2)]$. At fixed $N$, $\widetilde{S}^{\,\rm IPR}_{3} \simeq 2\widetilde{S}^{\,\rm IPR}_{2}$ to leading order in $1/D$, and the relative deviation reaches the plateau on the same logarithmic-in-$N$ depth scale as for $k=2$.}
\label{fig:ipr_U_qubits}
\end{figure}

\begin{funcbox}{brickwork\_average(B, d, N, t, bd)}
Contracts the full replica tensor network described in Sec.~\ref{sec:rtn:mps_evolution}. The routine builds the four-leg averaged gate $\widetilde{\mathcal{T}}^{(k)}$ from the basis \texttt{B} and local dimension \texttt{d}, initializes the row-MPS in the flat replica-spin state of Eq.~\eqref{eq:bc_init} corresponding to the product reference state $|0\rangle^{\otimes N}$, evolves it through $t$ brickwork layers using a TEBD-style two-site SVD update, and finally contracts the resulting MPS with the observable-dependent boundary amplitudes stored in \texttt{bd}. 
\end{funcbox}
The output of the contraction is the averaged inverse participation ratio $\mathbb{E}[\mathcal{I}^{(k)}(|\Psi_t\rangle)]$. 
Sweeping over a logarithmically spaced grid of system sizes $N$ and depths $t$ gives the data shown in Fig.~\ref{fig:ipr_U_qubits}: panels (a,b) correspond to $k=2$, while panels (c,d) correspond to $k=3$.

It is useful to visualize the same data in two complementary ways. Panels (a,c) show the bare quantity $\widetilde{S}_{(k)}^{\mathrm{IPR}}:=(1-k)^{-1}\log \mathbb{E}[\mathcal{I}^{(k)}]$, known as the annealed averaged participation~\cite{turkeshi2024hilbert}. Panels (b,d) instead show the deviation 
\begin{equation}
\begin{split}
\Delta \widetilde{S}_{(k)}^{\mathrm{IPR}}(t)
\;&=\;\widetilde{S}_{(k)}^{\mathrm{IPR}}(\infty) - \widetilde{S}_{(k)}^{\mathrm{IPR}}(t) 
\\
\widetilde{S}_{(k)}^{\mathrm{IPR}}(\infty)&=\frac{1}{1-k}\log \mathcal{I}^{(k)}_{\mathrm{Haar}} = \frac{1}{1-k}\log\left[\frac{k!\,(D)!}{(D + k - 1)!}\right] ,
\label{eq:DeltaI_rel}
\end{split}
\end{equation}
on a logarithmic vertical scale. This makes the convergence to the Haar plateau of Eq.~\eqref{eq:ipr_haar} directly visible.

The relative-deviation panels in Fig.~\ref{fig:ipr_U_qubits}(b,d) show a clean exponential approach to the Haar plateau. In this case, the RTN contraction reduces to the brickwork random-walk description of Refs.~\cite{dalzell2022random,turkeshi2024hilbert}, which gives
\begin{equation}
\Delta \tilde{S}_{k}^\mathrm{IPR}(t)
\;\propto\;
N e^{-\alpha_k t},
\qquad
\alpha_k > 0 ,
\label{eq:exp_decay}
\end{equation}
where the decay rate $\alpha_k$ depends on $k$ and on the local Hilbert-space dimension $d$, but not on $N$. The crossover from the early-time growth regime to the Haar plateau therefore occurs at the anticoncentration, or full Hilbert-space delocalization~\cite{Claeys_2025}, time
\begin{equation}
t_{\mathrm{AC}}(N)
\;\propto\;
{\log N}
\;+\;
O(1) ,
\label{eq:log_sat}
\end{equation}
which grows only logarithmically with system size. Thus, depths $t\gtrsim \log N$ are already sufficient to reproduce the $k$th moments of a Haar-random state to fixed accuracy~\cite{lami2024anticoncentration,magni2025anticoncentration,christopoulos2024universal}. 
This fact has practical impact, from  quantum-advantage benchmarking~\cite{boixo2018characterizing,arute2019quantum,dalzell2022random} to quantum learning via randomized measurements~\cite{schuster_random_2025,huang2020predicting,cioli2025approximate,Bertoni2024}.

The $k=3$ data in Fig.~\ref{fig:ipr_U_qubits}(c,d) are obtained by changing only the commutant basis, namely \texttt{SymmetricBasis(2)} to \texttt{SymmetricBasis(3)}. The basis size increases from $|\mathrm{S}_2|=2$ to $|\mathrm{S}_3|=6$, and the worst-case bulk bond dimension from $4$ to $36$. The IPR boundary remains site-independent, with $b_i(\sigma)=d$ for every $\sigma$, so the rest of the contraction is unchanged. We leave the explicit reproduction of panels (c,d) for $N\leq 64$ as an exercise~\footnote{Note that larger system sizes are easily attainable.}.

We conclude with a comment on self-averaging. In the computation above we evaluated the annealed participation entropy
$\widetilde{S}_{(k)}^{\mathrm{IPR}}=(1-k)^{-1}\log \mathbb{E}[\mathcal{I}^{(k)}]$.
In many applications, however, the quantity of interest is the corresponding quenched average
$\overline S_k^{\rm IPR}:=\mathbb{E}[(1-k)^{-1}\log \mathcal{I}^{(k)}]$.
At finite size these two quantities need not coincide, since the logarithm and the ensemble average do not commute.

In the thermodynamic scaling limit this distinction becomes immaterial for the present brickwork ensembles. The IPR is self-averaging: its relative variance
$\mathrm{Var}[\mathcal{I}^{(k)}]/\mathbb{E}[\mathcal{I}^{(k)}]^2$
decays exponentially with circuit depth, cf. Ref.~\cite{turkeshi2024hilbert}.
Consequently, fluctuations between circuit instances become negligible, and the replica tensor network, although formulated for annealed moments, also gives quantitative access to the typical, quenched-averaged participation entropy. The same mechanism applies to the local purity and to other observables of interest, and extends to the other circuit ensembles considered below.

\subsubsection{Worked example: local purity and the entanglement membrane}
\label{sec:applications_1_LP}
The power of the replica tensor network comes from its versatility.
For example, studying the propagation of the circuit-averaged local purity, and thus the annealed entanglement entropy, requires only changing the boundary vector.
In code, for half-system bipartition $A=\{1,\dots,N/2\}$, 
\begin{pybox}
bd_LP = rtn.RenyiPurityBoundary(B, d, range(1, N // 2 + 1))
val_LP = rtn.brickwork_average(B, d, N, t, bd_LP)
\end{pybox}
\noindent returns $\mathbb{E}[\mathcal{P}_2(t)] = \mathbb{E}[\Tr(\rho_A^2)]$.  
Here, following Ref.~\cite{zhou2,turkeshi2024hilbert} we can resolve the problem analytically, thus giving a strong benchmark for the implementation.

The idea comes from two considerations.
First, the explicit form of $\widetilde{\mathcal{T}}^{(2)}$ has the two ferromagnetic configurations $|\iota\iota\rrangle$ and $|s s\rrangle$ as absorbing fixed points (diagonal eigenvalue $1$), while the mixed configurations $|\iota s\rrangle$ and $|s\iota\rrangle$ are mapped back into the two ferromagnetic ones with weight $K_d = d/(d^2 + 1)$ each.  
In other words, the RTN corresponds to the partition function of a random walk of a \emph{single} domain wall separating a $|\iota\rrangle$-region from an $|s\rrangle$-region, and each layer-by-layer transition that moves the wall by one bond costs a factor $K_d$.

Second, the LP boundary fixes the wall at the bipartition edge $x = \ell$ at $t = 0$, while the flat initial-state condition projects it onto a uniform mixture of the two ferromagnetic ends; combining the two, the surviving world-lines $x(s)$ are paths of a one-dimensional simple random walk on $\{0, 1, \ldots, N\}$ with absorbing boundaries at $x = 0$ and $x = N$, starting at $x_0 = \ell$, and weighted by $(2 K_d)^s$ if the walker is still alive at depth $s$ or $(2 K_d)^{s_{\mathrm{abs}}}$ if it has been absorbed at depth $s_{\mathrm{abs}} \leq t$.  This is the so-called  \emph{entanglement membrane} picture~\cite{nahum2017quantum,zhou2019emergent}.  Summing over the first-passage-time distribution $u_{\ell, s}$ of the walk gives the closed form
\begin{equation}
\mathcal{P}_2(t; A)
\;=\;
(2 K_d)^t \sum_{s = t + 1}^{\infty} u_{\ell, s}
\;+\;
\sum_{s = 0}^{t} (2 K_d)^s\, u_{\ell, s} ,
\qquad
K_d = \frac{d}{d^2 + 1} ,
\label{eq:P2_exact}
\end{equation}
with the random-walk-absorption kernel is fixed by the boundary conditions
\begin{equation}
u_{z, t}
\;=\;
\sum_{\nu = 0}^{N/2 - 1} \frac{2}{N}\,
   \sin\!\Bigl(\tfrac{\pi (2\nu + 1)}{N}\Bigr)\,
   \cos^{\,t - 1}\!\Bigl(\tfrac{\pi (2\nu + 1)}{N}\Bigr)\,
   \sin\!\Bigl(\tfrac{\pi (2\nu + 1) z}{N}\Bigr) ,
\label{eq:uzt_exact}
\end{equation}
where $\ell = |A|$ and $K_d$ is the domain-wall step weight.  Eq.~\eqref{eq:P2_exact} is exact at any $d$.

\begin{figure}[t!]
\centering
\includegraphics[width=\textwidth]{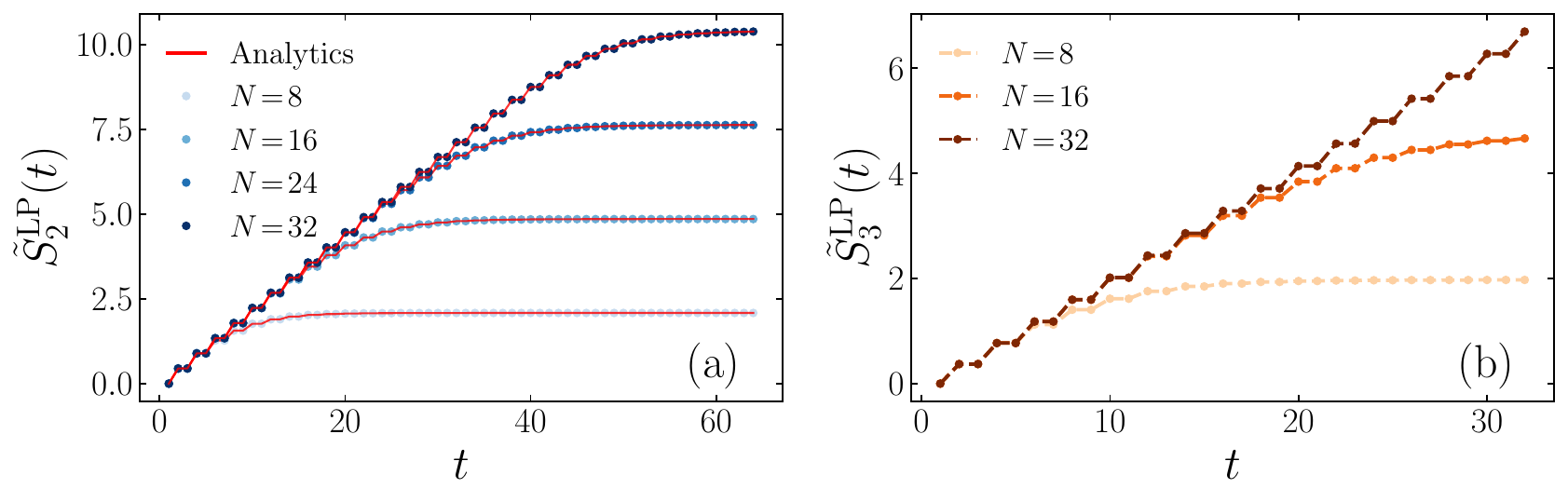}
\caption{Half-chain local purities on Haar-unitary qubit brickworks.  \emph{(a)} Annealed local-purity entropy $\widetilde{S}^{\,\rm LP}_{2}(t) = -\log \mathbb{E}[\mathcal{P}_2(t)]$ as a function of depth $t$ for $N \in \{8, 16, 24, 32\}$ qubits and $t \leq 64$ (Blues palette); markers are the RTN, solid lines the analytic random-walk closed form of~Eq.~\eqref{eq:P2_exact}.  \emph{(b)} The $k = 3$ analogue, $\widetilde{S}_{(3)}^{\mathrm{LP}}(t) = -\tfrac{1}{2}\log \mathbb{E}[\mathcal{P}_3(t)]$, computed by the RTN at $N \in \{8, 16, 32\}$ (Oranges palette); no closed form is known and the RTN is the only available estimator.}
\label{fig:purity_panel}
\end{figure}

We present the comparison with the replica tensor network numerics in Fig.~\ref{fig:purity_panel}(a) using the annealed entanglement entropy $\tilde{S}^\mathrm{LP}_2 = -\log \mathbb{E}[\mathcal{P}_2(t)]$.
At early times, $t \ll N$, the wall has not yet reached either end of the chain and Eq.~\eqref{eq:P2_exact} predicts $\log \mathcal{P}_2(t) \approx -v_d\,t$ with the so-called entanglement velocity $v_d = -\log(2 K_d)$ (for qubits $v_2 = \log(5/4) \approx 0.223$).  
At $t \sim N$ the wall reaches the absorbing boundary and the purity saturates to the Page value Eq.~\eqref{eq:page_value}.  The saturation timescale of the local purity is therefore $t_{\mathrm{ent}} \propto N$, \emph{ballistic} in the system size, in sharp contrast with the $\log N$ anticoncentration timescale~\eqref{eq:log_sat} of the IPR.  Physically, the IPR boundary~\eqref{eq:bc_ipr_flat} is a global, $\sigma$-independent constant per site that does not select any direction in $\mathrm{S}_k$-spin space, so the network can equilibrate before any large-scale correlation has been established; the LP boundary, instead, encodes a non-trivial bipartition through the cyclic-permutation factor on $B$, which forces a domain-wall configuration whose energetic cost scales as $|\partial A|$ and which can only saturate after light-cone propagation across the chain.

Repeating the computation at $k = 3$ requires only the basis change \texttt{SymmetricBasis(2) $\to$ SymmetricBasis(3)} and the corresponding LP boundary on six instead of two sigma-spins using the cyclic permutation $\eta=(123)$; the RTN itself is unchanged.  The closed-form bookkeeping does \emph{not} carry over: at $k = 3$ the transfer matrix has negative entries, so the corresponding stat-mech model has signed Boltzmann weights and the leading-domain-wall expansion is not a manifestly positive sum.  The RTN nevertheless evaluates the $k = 3$ purity directly, and panel~(b) of Fig.~\ref{fig:purity_panel} reports $\widetilde{S}_{(3)}^{\mathrm{LP}}(t) = -\tfrac{1}{2}\log \mathbb{E}[\Tr(\rho_A^3)]$ on the same dynamical grid as panel~(a).

In conclusion, the entanglement spreads linearly in time and saturates at a timescale $O(N)$ to the stationary Haar state. In contrast, the participation entropy spread much quicker, saturating in a timescale $O(\log N)$. This difference come from the boundary condition: the IPR requires a sum over all possible domain wall final configurations, from zero to $N$. 
A systematic analysis, beyond relevance for the present notes, is given in Ref.~\cite{dalzell2022random,christopoulos2024universal}.

\begin{exercise}
\label{sec:exercise_otoc_unitary}
\textbf{Out-of-time-ordered correlator (OTOC).}
Operator spreading, or scrambling, is probed by the \emph{out-of-time-ordered correlator}~\cite{roberts_chaos_2017}, 
\begin{equation}
    \mathcal{O}^{(2k)}(t) :=
\frac{1}{D}
\mathbb{E}\operatorname{tr}\!\left(
A_1 \mathbf{U}_t B_1 \mathbf{U}_t^\dagger
\cdots
A_k \mathbf{U}_t B_k \mathbf{U}_t^\dagger
\right)\;.
\end{equation}
The standard 4-point otoc is obtained for $k=2$. Assume $A_1=A_2=B_1=B_2=A$ is a local operator, e.g. a Pauli matrix $A=Z_i$. 
(a) Cast the computation of $\mathcal{O}^{(4)}(t)$ as a replica tensor network. What are the boundary and initial state?
(b) Compute the Haar value, assuming $\mathbf{U}_t$ is randomly sampled from $\mathrm{U}(D)$. What is its value for $A=Z_i$? 
(c) Implement the calculation in the replica tensor network for the average dynamics over the circuit. How fast does the OTOC converge to the Haar value? Which timescale is needed to approximate this value with finite precision $\varepsilon$?
\end{exercise}

\begin{exercise}
\label{sec:exercise_heisenberg}
\textbf{Faster scrambler.}
The two-site brick of these notes is fully Haar-random.  In experiments
the entangling gate is typically \emph{fixed} and only the single-qubit
unitaries are randomised.  This exercise probes that more realistic
ensemble.  Take
\begin{equation}
  U_{\rm elem} \;=\; (u_1 \otimes u_2)\,\exp\!\left[i \tfrac{\pi}{4}\,(XX + YY + ZZ)\right] ,
  \label{eq:elementary_gate}
\end{equation}
where $u_1, u_2$ are independent single-qubit Haar-random unitaries and
the central piece is the deterministic iSWAP, cf. Ref.~\cite{marko}.  
The single-qubit averaging gives a
\emph{closed-form} averaged 2-replica gate
$\widetilde{\mathcal{T}}^{(k)}_{\rm elem}$, at the cost of integrating
$u_1, u_2$ analytically while leaving the Heisenberg piece untouched.
(a) Sandwich $k$ copies of
$U_{\rm elem} \otimes U_{\rm elem}^*$ between $|\pi_1\rrangle|\pi_2\rrangle$
at the bottom and $\bbra{\sigma_1}\bbra{\sigma_2}$ at the top.
Integrate over $u_1, u_2$ using the standard
$\mathbb{E}[u^{\otimes k}\otimes u^{*\otimes k}]$ formulae of
\S\ref{sec:rtn:average}; the result is an $|\mathrm{S}_k|^2 \times |\mathrm{S}_k|^2$
tensor.  Write it down in the basis of $\mathrm{S}_k$ for $k=2$ and $k=3$. 
(b) Implement the simulation of the IPR for $k=2,3$ within this circuit. Does it converge faster or slower in circuit depth compared to the random Haar circuit?
\end{exercise}

\subsection{Irrep-reduced commutant basis*}
\label{sec:irrep_reduction}
On the $(\mathbb{C}^d)^{\otimes k}$ representation, 
the raw symmetric-group basis of size $|\mathrm{S}_k| = k!$ is, in general,
\emph{over-complete}: whenever $d < k$ the Gram matrix $G_{\sigma \tau}
= d^{\#(\sigma^{-1}\tau)}$ has a non-trivial kernel, and several
different permutation operators $P_\sigma \in
\mathrm{End}((\mathbb{C}^d)^{\otimes k})$ are linearly dependent.
The effective commutant therefore lives in a smaller subspace of
dimension $d_{\mathrm{red}}(k, d) \leq k!$.  Contracting the replica
tensor network on this reduced subspace is an exact rewriting of the
full calculation, no information is lost, and the "spin"
dimension of the tensor network drops from $|\mathrm{S}_k|^2$ to
$d_{\mathrm{red}}^2$, a sizeable gain once $k \geq 4$.

The canonical example is $k = 4$ qubits:

\begin{equation}
|\mathrm{S}_4|  \;=\;  24
  \quad \xrightarrow{d = 2}  \quad
  d_{\mathrm{red}}(4, 2)  \;=\;  C_4 \;=\; 14,
\label{eq:irrep_24to14}
\end{equation}
where $C_k = (2k)!/[k!(k+1)!]$ is the $k$th Catalan number.
The bulk transfer matrix drops from  $24^2 = 576$ to $14^2 = 196$, an almost $3\times$
speed-up.  The same mechanism gives $120 \to 42$ at $k = 5$ qubits,
and $6 \to 5$ at $k = 3$ qubits.  For $d \geq k$ the reduction is
trivial: by Schur--Weyl duality the $k!$ permutation operators are
already linearly independent on $(\mathbb{C}^d)^{\otimes k}$ and
$d_{\mathrm{red}} = k!$.

Following Ref.~\cite{Braccia2024,turkeshi2025magic}, the projector onto the
irreducible subspace is built from the (real symmetric) Gram matrix
$G$ alone:

\begin{equation}
\mathsf{Id}  \;=\;  G^{+} G,
\qquad
\mathsf{Id}  \;=\;  \sum_{\alpha = 1}^{d_{\mathrm{red}}} |v_\alpha\rrangle\!\llangle v_\alpha|
\quad \text{(eigenvalue 1)},
\label{eq:irrep_identity}
\end{equation}
where $G^{+}$ is the Moore--Penrose pseudo-inverse and
$\{|v_\alpha\rrangle\}$ is any orthonormal basis of the eigenspace of
$\mathsf{Id}$ with eigenvalue $+1$ (equivalently, of the column space
of $G$).  Stacking these basis vectors into a
$(d_{\mathrm{red}} \times |\mathrm{S}_k|)$ matrix $P$ with $P P^{T} =
\mathbb{I}_{d_{\mathrm{red}}}$, the reduced bulk transfer matrix and
boundary vector are
\begin{equation}
\Phi_{t,\mathrm{red}}^{(k)}  \;=\;  (P \otimes P)\,
  \Phi_t^{(k)}\, (P \otimes P)^{T}, \qquad
\llangle \Omega^{{\Lambda}}_k |_{\mathrm{red}}
  \;=\;  \llangle \Omega^{{\Lambda}}_k|\, P^{T},\qquad |\mathfrak{I}^{(k)}\rrangle_\mathrm{red} = P |\mathfrak{I}^{(k)}\rrangle\;,
\label{eq:reduced_objects}
\end{equation}
and the RTN contraction runs on the reduced site space of dimension
$d_{\mathrm{red}}$ with the same code path as the raw basis.  The value
of every scalar observable is preserved identically, 

\begin{equation}
\llangle \Omega^{\Lambda}_k | \Phi_t^{(k)} |\mathfrak{I}^{(k)}\rrangle
  \;=\;
\llangle \Omega^{\Lambda}_k |_{\mathrm{red}}\Phi_{t,\mathrm{red}}^{(k)}
        |\mathfrak{I}^{(k)}\rrangle_{\mathrm{red}} \;,
\label{eq:irrep_invariance}
\end{equation}
because $P^T P$ is the orthogonal projector onto the physically
meaningful subspace.  Equation~\eqref{eq:irrep_invariance}
is the practical statement of the reduction: the spin space defining the tensor network is smaller, and the observable it returns is
identical to the one the full-basis contraction would have
produced. 
Table~\ref{tab:irrep_red}
collects the effective dimensions for the cases that arise in these
lectures.  For $d \geq k$ (the $\llap{\raisebox{-0.15em}{$\cdot$}}$-marked
rows) the reduction is trivial.

\begin{table}[h]
\centering
\begin{tabular}{c c c c c}
$k$ & $d$ & $|\mathrm{S}_k|$ & $d_{\mathrm{red}}(k, d)$ & $|\mathrm{S}_k|^2 / d_{\mathrm{red}}^2$ \\
\hline
$2$ & $2$ & $2$   & $2$    & $1.00\times$ \\
$3$ & $2$ & $6$   & $5$    & $1.44\times$ \\
$3$ & $3$ & $6$   & $6\,{}^{\llap{\raisebox{-0.15em}{$\cdot$}}}$  & $1.00\times$ \\
$4$ & $2$ & $24$  & $\mathbf{14}$ ($C_4$) & $\mathbf{2.94\times}$ \\
$4$ & $3$ & $24$  & $23$   & $1.09\times$ \\
$4$ & $4$ & $24$  & $24\,{}^{\llap{\raisebox{-0.15em}{$\cdot$}}}$ & $1.00\times$ \\
$5$ & $2$ & $120$ & $\mathbf{42}$ ($C_5$) & $\mathbf{8.16\times}$ \\
\end{tabular}
\caption{Effective commutant dimension $d_{\mathrm{red}}(k, d)$ on
$(\mathbb{C}^d)^{\otimes k}$ after projection onto the column space
of the Gram matrix.  For qubits the reduction saturates the Catalan
numbers $C_k$; for $d \geq k$ it is trivial by Schur--Weyl duality.
The last column gives the speedup form the trivial implementation.
}
\label{tab:irrep_red}
\end{table}

The reduction is exposed by a handful of primitives.  Given any
concrete list of
commutant operators the reader can build the reduced objects in three
lines:

\begin{pybox}
import replicatn as rtn
# Haar unitary, k = 4 qubits: 24 -> 14
sk      = rtn.symmetric_group_matrices(4, 2)
G       = rtn.irrep_gram(sk)
W       = rtn.haar_unitary_bulk_gate(sk, d=2)    # full 576x576
P, dred = rtn.irrep_projector(G)                 # dred = 14
Wred    = rtn.irrep_reduce_gate(W, P)            # 196x196
bnd_red = rtn.irrep_reduce_boundary(rtn.ipr_boundary(sk), P)
\end{pybox}
The reduced pair $(W_{\mathrm{red}},\, b_{\mathrm{red}})$ plugs into
\texttt{brickwork\_rtn\_average} with the same interface as the raw
objects; the only change the reader observes is a smaller bond
dimension and a correspondingly faster contraction.

At the level of representation theory,
the procedure is an instance of Schur--Weyl duality on a finite
representation: the commutant of $\mathrm{U}(d)$ acting diagonally on
$k$ tensor factors is spanned by $\mathbb{C}[\mathrm{S}_k]$, but on a
representation of finite dimension this span is typically proper.  For
qubits, the span is exactly the Temperley--Lieb algebra
$\mathsf{TL}_k(\delta = 2)$~\cite{turkeshi2025magic}; the
positive-eigenvalue subspace of $\mathsf{Id} = G^+ G$ is
$\mathsf{TL}_k$ itself, and its dimension is the Catalan number $C_k$.
For qutrits one is building, \emph{mutatis mutandis}, the analogous
quotient of $\mathbb{C}[\mathrm{S}_k]$ with respect to the kernel of the
restriction map to $\mathrm{End}((\mathbb{C}^3)^{\otimes k})$.

\FloatBarrier
\section{Noisy random circuits}
\label{sec:noisy_circuits}

\label{sec:noisy}

\subsection{Noise models}
\label{sec:noisy:model}

This section studies how noise affects the dynamics of information in
random quantum circuits.  We start with the simplest case of an
identical depolarising channel acting on every site after every
layer, and then turn to the asymmetric situation in which only one
copy of the circuit is hit by noise while the other is the ideal
quantum-state output, the natural setting for benchmarking
quantum-classical correlations~\cite{boixo2018characterizing,
arute2019quantum, sauliere2026errorcorrectiontransitionsfinitedepthquantum}.

We focus on the local single-qudit \emph{depolarising channel}, which
on a $d$-dimensional qudit reads
\begin{equation}
\mathcal{N}_p(\rho)
\;=\;
(1 - p)\,\rho \;+\; p\,\frac{\mathbb{I}_d}{d}\,\mathrm{tr}(\rho) ,
\qquad p \in [0, 1] ,
\label{eq:depolarising_local}
\end{equation}
and the noisy circuit applies the tensor product
$\mathcal{N}_p^{\otimes N}$ to the global state after every brickwork
layer.  Concretely, every site is independently replaced by the
local maximally-mixed state with probability $p$, and left untouched
with probability $1 - p$.  At $p = 0$ this is the identity (clean
evolution); at $p = 1$ each layer resets the chain to the maximally
mixed state $\mathbb{I}_d/d$ on every site.  In the doubled-space
representation $\mathcal{N}_p$ has the Choi matrix
\begin{equation}
\mathcal{N}_p
\;=\;
(1 - p)\,\mathbb{I}^{\otimes 2}
\;+\;
\frac{p}{d}\,|e\rrangle\llangle e| ,
\qquad
|e\rrangle = \sum_{a = 0}^{d - 1}\,|a, a\rangle / \sqrt{d} ,
\label{eq:depolarising_choi}
\end{equation}
where $|e\rrangle$ is the normalised maximally entangled state on the
doubled single-qudit space.

\begin{funcbox}{depolarising\_choi(d, p)}
Returns the $d^2\!\times\!d^2$ Choi matrix
of~Eq.~\eqref{eq:depolarising_choi}.  Plugs directly into
\texttt{noisy\_brickwork\_rtn\_average(B, d, N, t, bd, channels)} as
one entry of the per-replica \texttt{channels} list of length $k$.
\end{funcbox}

\subsection{Incorporating noise into the replica tensor network}
\label{sec:noisy:incorporating}

The noise channel is local and deterministic: in the $k$-replica
formalism, the action of $\mathcal{N}_p$ on $k$ copies of a single
qudit factorises as the per-site tensor product
$\mathcal{N}_1\otimes  \mathcal{N}_2\otimes \cdots \otimes \mathcal{N}_k $, where potentially we can have different noise acting on different copies of the system. 
We will consider a circuit where noise acts locally on each qudit that was evolved with a unitary gate. In other words, the action of each unitary gate $U_{i,i+1}A U^\dagger_{i,i+1}$ is replaced with $(\mathcal{N}_i\circ\mathcal{N}_{i+1})(U_{i,i+1}A U^\dagger_{i,i+1})$.

We now detail the derivation of the noisy transfer matrix in the way that actually
gets implemented in the code, see e.g.~\cite{sauliere2026errorcorrectiontransitionsfinitedepthquantum,arman}.

Applying noise channel on qudits after the unitary gates leads to the replacement
\begin{equation}
  \mathcal{T}_{i,i+1}\to  \left(\bigotimes_{i=1}^k\mathcal{N}_i\right)\mathcal{T}_{i,i+1} =
\sum_{\sigma, \pi} \mathrm{Wg}_{\pi, \sigma} \left(\bigotimes_{i=1}^k\mathcal{N}_i\right)|\sigma\rrangle \llangle \pi |  \;.
\label{eq:ziopino}
\end{equation}
When contracting with the next gate, the action of $\mathcal{N}^{\otimes k}$ leads to the \emph{noisy Gram matrix}
\begin{equation}
[\tilde{G}(q; \mathcal{N}_1, \ldots, \mathcal{N}_k)]_{\pi, \sigma}
   :=  \llangle \pi \bigl| \mathcal{N}_1 \otimes \mathcal{N}_2 \otimes \cdots \otimes \mathcal{N}_k \bigr| \sigma \rrangle,
\label{eq:noisy_overlap_alt}
\end{equation}
where $\mathcal{N}_a$ is the Choi superoperator of the channel on
replica $a$ (a $d^2 \times d^2$ matrix) and the inner product is over
the single-site $d^{2k}$-dim doubled replica space.  For \emph{identity}
channels, $\tilde{G}$ reduces to the ordinary Gram matrix
$G_{\pi, \sigma}(d) = d^{\#(\pi^{-1}\sigma)}$, so
$\tilde{G}$ is the precise minimal generalization that captures the
noise.

The power of this formulation is that \emph{changing the ensemble or
the noise model changes only the entries of $\tilde{G}$}: the bulk
structure of the tensor network, how layers chain together, how the
boundaries contract, is untouched.  Setting $\mathcal{N}_1 =
\mathbb{I}$ and $\mathcal{N}_2 = \mathcal{N}$ for $k = 2$ yields the
\emph{XEB} structure (one replica clean representing the classical computation, one noisy accounting for the quantum computation); setting
$\mathcal{N}_a = \mathcal{N}$ for every replica yields the symmetric
$k$-copy noisy collision probability.

In essence, to assemble $\widetilde G$ one taks the $d^{2k}$-dim
Kronecker product of the $k$ Choi matrices and contract against the
permutation kets. 
This strategy works even when the circuit is partially random and partially deterministic: one relevant example is a circuit with local random on-site Haar gates alternated with deterministic ones~\cite{marko,suzuki2025globalrandomnessrandomlocal}.

In summary, the noisy contraction differs from the clean
Algorithm~\ref{alg:rtn_brickwork} by simply changing who are the building gates $\tilde{\mathcal{T}}$ that now incorporate the noisy Gram matrices.  

\begin{funcbox}{noisy\_brickwork\_rtn\_average(B, d, N, t, bd, channels)}
Same as \texttt{brickwork\_rtn\_average} but with the inclusion of the noise acting on
each qudit.  The per-replica Choi matrices live in
\texttt{channels} (length $k$); for the depolarising case used in this
section, every entry is built with \texttt{depolarising\_choi(d, p)}.
\end{funcbox}

\subsection{Worked example: relative entropy of coherence under symmetric noise}
\label{sec:noisy:rel_entropy_coh}

\label{sec:coherence_relative}

Coherence of a quantum state $\rho$ relative to a fixed reference
basis (the computational basis $\{|\pmb{x}\rangle\}$ throughout these notes)
is the property of having non-zero off-diagonal entries in that
basis.  When the system is mixed, a standard quantitative diagnostic, cf. e.g.
Refs.~\cite{baumgratz_quantifying_2014,streltsov_colloquium_2017,kelly,aditya,aditya2025growthspreadingquantumresources,aditya2025mpembaeffectsquantumcomplexity},
is the \emph{relative entropy of coherence}
\begin{equation}
C^{(2)}_{\mathrm{rel}}(\rho)
\;=\;
\log\,\frac{\Tr(\rho^2)}{\Tr(\rho^2_{\mathrm{diag}})}
\;=\;
S^{(2)}(\rho_{\mathrm{diag}}) \,-\, S^{(2)}(\rho) ,
\label{eq:Crel2}
\end{equation}
where $\rho_{\mathrm{diag}} = \sum_{\pmb{x}} \langle \pmb{x} | \rho | \pmb{x}\rangle\,|\pmb{x}\rangle\langle \pmb{x}|$
is the fully dephased image of $\rho$ in the chosen basis, and
$S^{(2)}(\sigma) = -\log \Tr(\sigma^2)$ is the R\'enyi-2
entropy.  $C^{(2)}_{\mathrm{rel}}$ vanishes when $\rho$ is already
diagonal in $\{|\pmb{x}\rangle\}$ (no off-diagonal weight, no coherence) and
is bounded above by $\log D$ (saturated when $\rho$ is pure and maximally delocalized).  It
is the simplest non-linear diagnostic that is sensitive to the
\emph{quantumness} of the output bit-string distribution and not just
to its flatness.

For pure states evolving unitarily, $\Tr(\rho^2) = 1$ and so
$C^{(2)}_{\mathrm{rel}}(|\Psi_t\rangle) = -\log \mathcal{I}^{(2)}(|\Psi_t\rangle)$
is just the collision probability (up to a change in logarithmic base).  
After one single layer, the system develops extensive coherences due to the unitary gates. 
The noise starts to kick in and increases the thermal entropy, making the relative entropy of coherences to decrease to zero as $\rho_t \to \mathbb{I}/D$.  

\begin{figure}[t!]
\centering
\includegraphics[width=\textwidth]{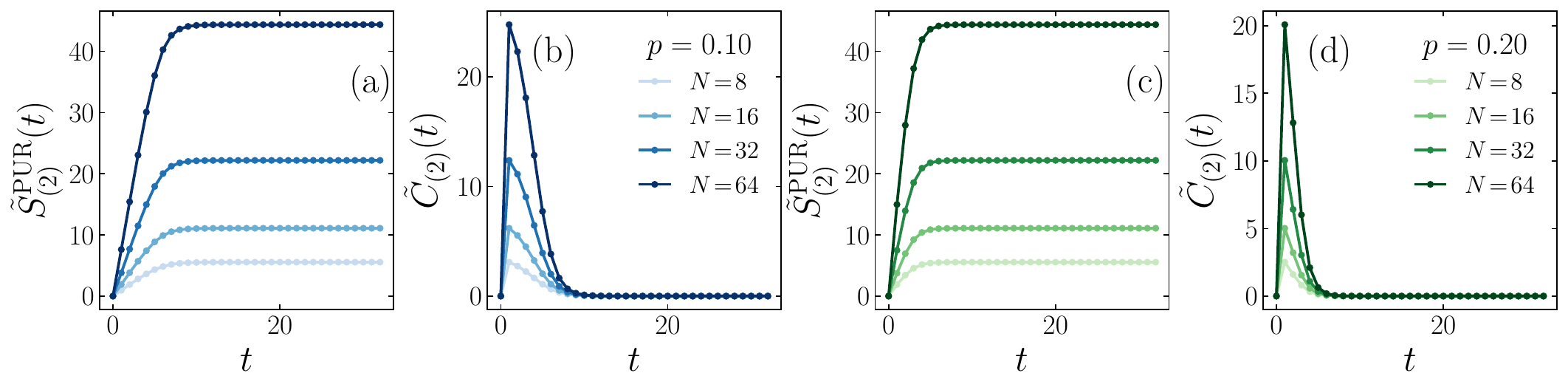}
\caption{Noisy random Haar-unitary brickwork on qubits, at every
integer depth and for $N \in \{8, 16, 32, 64\}$.  Panels (a) and
(c) plot the negative log purity $-\log \mathbb{E}[\Tr\,
\rho_t^2]$ at $p = 0.10$ and $p = 0.20$ respectively;
panels (b) and (d) plot the relative entropy of
coherence $\tilde{C}^{(2)}_{\mathrm{rel}}(\rho_t)$
at the same two noise rates.  The coherence rises linearly with
depth, peaks at the sampling transition, and decays to zero as
$\rho_t \to \mathbb{I}/D$ becomes classical.}
\label{fig:coherence_noisy}
\end{figure}

Both $\mathbb{E}[\Tr(\rho_t^2)]$ and
$\mathbb{E}[\Tr(\rho_{t, \mathrm{diag}}^2)]$ are
$k = 2$ doubled-replica observables.  Crucially, they share the
\emph{same} noisy bulk transfer matrix, and
differ only by the top boundary:
\begin{equation}
\tilde{C}^{(2)}_{\mathrm{rel}}(\rho_t)
\;:=\;
\log\,
\frac{\llangle \Omega^{\mathrm{full}}_2 \,|\,
   (\Phi_t^{(2)})_{\mathrm{noisy}} \,|\,\rho_0^{\otimes 2}\rrangle}
{\llangle \Omega^{\textup{IPR}}_2 \,|\,
   (\Phi_t^{(2)} )_{\mathrm{noisy}}\,|\,\rho_0^{\otimes 2}\rrangle} ,
\label{eq:Crel2_RTN}
\end{equation}
with $|\Omega^{\textup{IPR}}_2\rrangle$ the IPR boundary
of Eq.~\eqref{eq:boundary_IPR} ($b_i(\sigma) = d$ on every site) and
$|\Omega^{\mathrm{full}}_2\rrangle = \bigotimes_{i = 1}^N |s\rrangle_i$ the
full-chain swap boundary (the LP boundary of~\eqref{eq:boundary_LP_kets}
specialised to $A=\{1,\dots,N\}$, hence the swap on every site).  The two contractions are produced by a single
call to the function \texttt{noisy\_brickwork\_rtn\_average} with two different
boundary objects: \texttt{IPRBoundary(B, d)} for the denominator and
\texttt{RenyiPurityBoundary(B, d, 1:N)} for the numerator.

Fig.~\ref{fig:coherence_noisy} shows the noisy purity
$\widetilde{S}_{(2)}^{\mathrm{PUR}}(t):=-\log\mathbb{E}[\Tr\,\rho_t^2]$ and the relative entropy of
coherence $\tilde{C}^{(2)}_{\mathrm{rel}}(\rho_t)$ on $N \in \{8, 16, 32,
64\}$ qubits at two values of the local depolarising rate,
$p = 0.10$ and $p = 0.20$.  
After the initial peak, the coherent evolution decays back to zero as the noisy steady state
$\rho_\infty = \mathbb{I}_D/D$ takes over with time.

\FloatBarrier
\FloatBarrier
\subsection{Worked example: coherent information under symmetric noise}
\label{sec:noisy:coherent_info}

\label{sec:coherent_info}

A second physically meaningful observable accessible to the noisy RTN
is the \emph{coherent information} of the noisy encoding
channel~\cite{sauliere2026errorcorrectiontransitionsfinitedepthquantum}, the natural
error-resilient diagnostic~\cite{turkeshi2024errorresilience,lovas2024quantum} for the random circuits we are
discussing.  We restrict ourselves to local depolarising noise.

The construction takes $K$ logical qudits and entangles each with one
qudit of a clean reference system $R$ of $K$ qudits via $K$ Bell
pairs.  The remaining $N - K$ qudits of the physical system $B$ are
prepared in $|0\rangle^{\otimes (N - K)}$, and the noisy brickwork
acts on the entire $B$ (the reference $R$ is left untouched).  In
symbols, the initial state is
\begin{equation}
\rho_{RB}(0)
\;=\;
|\psi_+\rangle\langle \psi_+|_{R, B_{1:K}}
\;\otimes\;
\bigl(|0\rangle\langle 0|\bigr)^{\otimes (N - K)}_{B_{K+1:N}} ,
\qquad
|\psi_+\rangle = d^{-K/2}\,\sum_{\mathbf j}\,|\mathbf j\rangle_R\,|\mathbf j\rangle_B ,
\label{eq:codeRB_init}
\end{equation}
and the encoded state at depth $t$ is
$\rho_{RB}(t) = (I_R \circ \mathcal{U}_t)[\rho_{RB}(0)]$, where $I_R$ is the identity channel on the reference, and $\mathcal{U}_t$ is the noisy random circuit.
We assume the noise acts \emph{within} the encoding circuit, i.e. a
single-qudit depolarising channel $\mathcal{N}_p$ is applied to every
site of $B$ after every unitary gate layer.  Diagrammatically, with
time running upward,
\begin{equation}
\rho_{RB}(t)
\;=\;
\begin{tikzpicture}[baseline=(current bounding box.center), scale=0.55]
\definecolor{skyromeblue}{RGB}{86,180,233}
\definecolor{darkgraylines}{RGB}{55,55,55}
\definecolor{noisegrey}{RGB}{145,145,150}

\def\nqubits{6}
\def\ndepth{4}
\def\klog{3}
\def\gateborder{0.9pt}

\def\hookY{-2.0}

\draw[decorate, decoration={brace, amplitude=4pt}, thick, darkgraylines]
  (-\klog-0.4, \ndepth+1.05) --
  node[above=2pt]{\footnotesize $R$}
  (-0.6, \ndepth+1.05);
\draw[decorate, decoration={brace, amplitude=4pt}, thick, darkgraylines]
  (0.6, \ndepth+1.05) --
  node[above=2pt]{\footnotesize $B$}
  (\nqubits+0.4, \ndepth+1.05);

\draw[->, thick, darkgraylines]
  (\nqubits+1.0, 0) -- node[right=3pt]{\footnotesize $t$}
  (\nqubits+1.0, \ndepth+0.4);

\foreach \x in {1,...,\nqubits}{
  \draw[thick, darkgraylines] (\x, 0) -- (\x, \ndepth+0.85);
}

\foreach \i in {1,...,\klog}{
  \pgfmathsetmacro{\rxtop}{-\i + 0.5 - 0*\klog}     
  \pgfmathsetmacro{\bxbot}{\klog - \i + 1}          
  \pgfmathsetmacro{\hooklow}{\hookY - 0.3*\i}       
  \draw[thick, darkgraylines, rounded corners=8pt]
    (\rxtop, \ndepth+0.85) -- (\rxtop, \hooklow)
                           -- (\bxbot, \hooklow)
                           -- (\bxbot, 0);
}

\foreach \x in {\numexpr\klog+1\relax,...,\nqubits}{
  \draw[thick, darkgraylines] (\x, -0.05) circle (3pt);
  \node[darkgraylines, below=2pt] at (\x, -0.15) {\scriptsize $|0\rangle$};
}

\foreach \d in {0,...,\numexpr\ndepth-1\relax}{
  \pgfmathsetmacro{\y}{\d + 0.5}
  \ifodd\d
    \foreach \x in {2, 4}{
      \draw[line width=\gateborder, draw=skyromeblue!70!black,
            fill=skyromeblue, rounded corners=2pt]
        (\x-0.2, \y-0.25) rectangle (\x+1.2, \y+0.25);
    }
  \else
    \foreach \x in {1, 3, 5}{
      \draw[line width=\gateborder, draw=skyromeblue!70!black,
            fill=skyromeblue, rounded corners=2pt]
        (\x-0.2, \y-0.25) rectangle (\x+1.2, \y+0.25);
    }
  \fi
}

\foreach \d in {0,...,\numexpr\ndepth-1\relax}{
  \pgfmathsetmacro{\y}{\d + 1.0}
  \ifodd\d
    \foreach \x in {2,...,\numexpr\nqubits-1\relax}{
      \draw[line width=0.4pt, draw=darkgraylines, fill=noisegrey,
            shift={(\x, \y)}, rotate=30]
        (0:0.16) -- (60:0.16) -- (120:0.16) -- (180:0.16) --
        (240:0.16) -- (300:0.16) -- cycle;
    }
  \else
    \foreach \x in {1,...,\nqubits}{
      \draw[line width=0.4pt, draw=darkgraylines, fill=noisegrey,
            shift={(\x, \y)}, rotate=30]
        (0:0.16) -- (60:0.16) -- (120:0.16) -- (180:0.16) --
        (240:0.16) -- (300:0.16) -- cycle;
    }
  \fi
}

\draw[line width=0.4pt, draw=darkgraylines, fill=noisegrey,
      shift={(\nqubits+2.2, 0.5)}, rotate=30]
  (0:0.16) -- (60:0.16) -- (120:0.16) -- (180:0.16) --
  (240:0.16) -- (300:0.16) -- cycle;
\node[darkgraylines, right] at (\nqubits+2.4, 0.5)
      {\footnotesize $= \mathcal{N}_p$};

\end{tikzpicture}\;.
\label{eq:codeRB_circuit}
\end{equation}

The natural figure of merit is the \emph{coherent information}
$I_c(\mathcal{U}) = \mathrm{S}_{(2)}(\rho_B(t)) - \mathrm{S}_{(2)}(\rho_{RB}(t))$~\cite{Gullans2020,Colmenarez2024,Qian2025}, normalised by~$K$:
$I_c/K = 1$ signals perfect decodability, $I_c/K < 1$ irreversible
loss of logical information.  
We consider the annealed averaged circuit
\begin{equation}
I_c^{(2)}(\mathcal{U})
\;=\;
\widetilde{S}_{(2)}^{\mathrm{PUR}}(B) - \widetilde{S}_{(2)}^{\mathrm{PUR}}(R\cup B)
\label{eq:Ic2}
\end{equation}
which lifts to a $k = 2$ replica observable with the same noisy bulk
for both terms; only the top boundary changes.

\begin{figure}[!htbp]
\centering
\includegraphics[width=\linewidth]{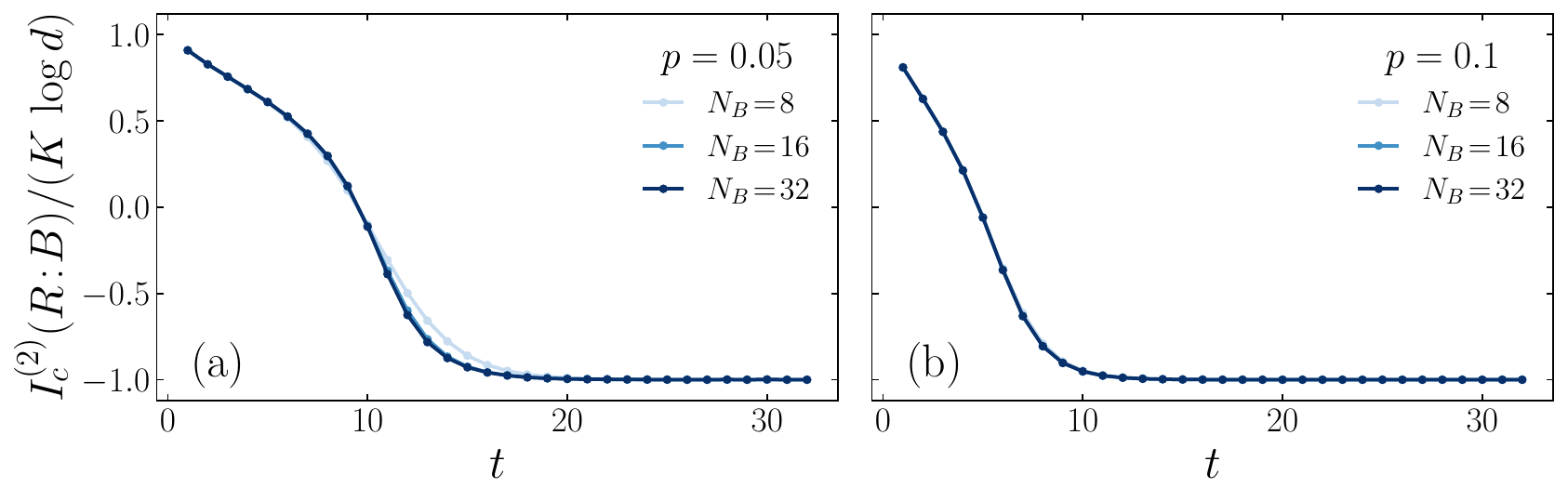}
\caption{R\'enyi-2 coherent information
$I_c^{(2)}(R\!:\!B)/(K\,\log d)$ of the noisy Haar-random encoder on
$N_B$ qubits as a
function of circuit depth~$t$, for two values of the single-qubit
depolarising strength $p \in \{0.05, 0.10\}$ at fixed
$K = 1$ reference qubit.  RTN prediction at
$N_B \in \{8, 16, 32\}$.  As $N_B$ grows the curves approach
a sharp crossover at a critical depth $t_c(p)$, the
error-correction transition between the recoverable phase
$I_c^{(2)} \to +1$ and the decohered phase $I_c^{(2)} \to -1$.}
\label{fig:coherent_info}
\end{figure}

Fig.~\ref{fig:coherent_info} reports $I_c^{(2)}/K$ as a function of
depth for $d = 2$, $K = 1$, $N \in \{8, 16, 32, 64\}$ at two
depolarising rates $p = 0.05$ (panel a) and $p = 0.10$
(panel b).  The curves display the error-correction transition
cleanly: at fixed $p$ they collapse onto a sharp crossover at a
critical depth $t_c(p)$, separating a low-depth regime where
the logical qudit is still recoverable ($I_c^{(2)}/K \approx 1$)
from a high-depth regime where decoherence has destroyed the logical
information ($I_c^{(2)}/K \approx -1$).  

\subsection{Worked example: the linear cross-entropy benchmark with asymmetric noise}
\label{sec:noisy:xeb}

The relative entropy of coherence diagnostic of
\S\ref{sec:noisy:rel_entropy_coh} placed the same noise on both
replicas.  In experiments, however, one almost always has access to
two \emph{distinct} ensembles of states: the actual outputs of a
quantum processor running the noisy circuit, and the perfect outputs
of the same circuit simulated on a classical computer.  The
asymmetric $k = 2$ replica observable that compares them is the
\emph{linear cross-entropy benchmark}
\begin{equation}
\chi_{\mathrm{XEB}}(t)
\;=\;
d^N\,\mathbb{E}\left[\sum_{\pmb{x} \in \mathbb{Z}_d^N}
   p_{\mathrm{cl}}(\pmb{x})\,p_{\mathrm{n}}(\pmb{x})\right] {\;-\; 1} ,
\label{eq:xeb_def}
\end{equation}
where $p_{\mathrm{cl}}(\pmb{x}) = |\langle \pmb{x} | U |\Psi_0\rangle|^2$ is the
ideal output distribution, whereas the probability distribution
$p_{\mathrm{n}}(\pmb{x}) = \langle \pmb{x} | \mathcal{U}_t(|\Psi_0\rangle\langle \Psi_0|) | \pmb{x}\rangle$ is the
output from the noisy device.  The observable $\chi_{\mathrm{XEB}}$ plays a central
role in benchmarking quantum-advantage regimes in current quantum devices, cf. Ref.~\cite{arute2019quantum,morvan2024phaseTransition}. 
The $-1$ subtraction in Eq.~\eqref{eq:xeb_def} is the
standard normalisation that turns $\chi_{\mathrm{XEB}}$ into a fidelity-like
diagnostic: a perfect quantum simulation ($p_{\mathrm{n}} = p_{\mathrm{cl}}$, no
device error) gives $\chi_{\mathrm{XEB}} \to 1$ at depths $t \gtrsim \log N$
where the Haar limit $d^N\,\mathbb{E}[\sum_{\pmb{x}} p^2] = 2 d^N / (d^N + 1)$ saturates,
while uniformly-random outputs ($p_{\mathrm{n}} \equiv d^{-N}$, fully decohered
device) give $\chi_{\mathrm{XEB}} \to 0$.  Empirically, $\chi_{\mathrm{XEB}}(t)$ is
thus an exponentially-decaying proxy for the running fidelity of the
noisy circuit relative to its ideal classical simulation.

\begin{figure}[t!]
\centering
\includegraphics[width=\linewidth]{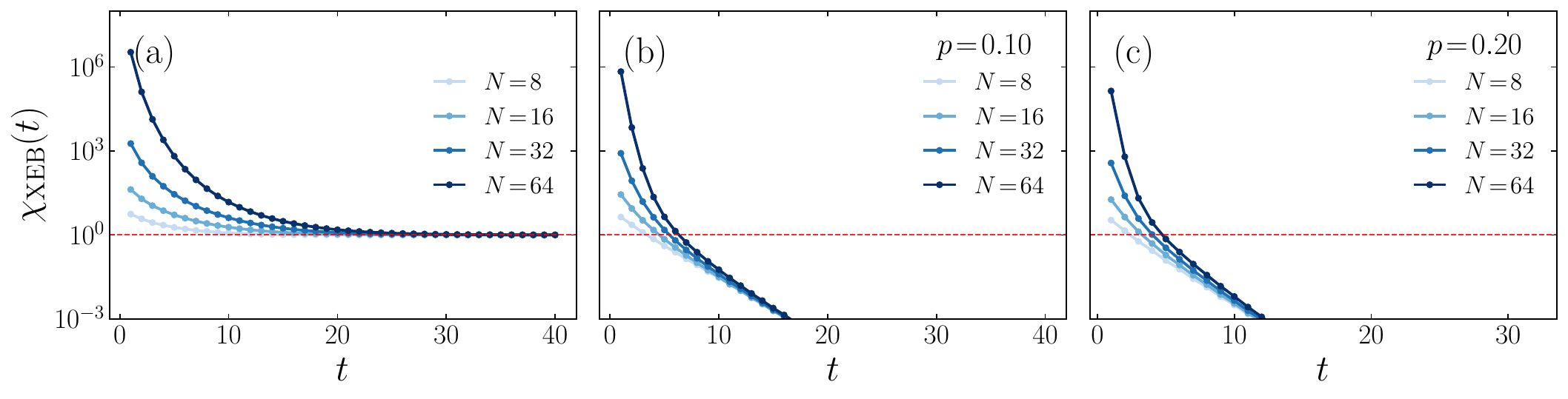}
\caption{Linear cross-entropy benchmark
$\chi_{\mathrm{XEB}}(t)$~Eq.~\eqref{eq:xeb_def} for a Haar-random
brickwork on $N \in \{8, 16, 32, 64\}$ qubits up to depth $t = 40$.
\emph{(a)} Clean reference (no device error), saturating
at the Haar fidelity plateau $\chi_{\mathrm{XEB}} \to (D - 1) / (D + 1)
\to 1$ at depths $t \gtrsim \log N$.  \emph{(b)} Incoherent
depolarising noise on the device replica, rate $p = 0.10$.
\emph{(c)} Same with $p = 0.20$.  In both noisy panels,
$\chi_{\mathrm{XEB}}$ decays exponentially towards the classical
(no-quantum-advantage) regime $\chi_{\mathrm{XEB}} \to 0$.  Log-scale
$y$-axis.}
\label{fig:xeb_k2}
\end{figure}

At the level of the RTN, $\chi_{\mathrm{XEB}}$ is the asymmetric
noisy-overlap matrix~Eq.~\eqref{eq:ziopino} with
$\mathcal{N}_1 = \mathbb{I}$ (clean simulation in one replica)
and $\mathcal{N}_2$ the channel implemented on the device
(noisy second replica).  In the companion package, this is one
\texttt{noisy\_brickwork\_rtn\_average} call with
$k = 2$ and a \texttt{channels} list whose first entry is the
identity Choi:
\begin{pybox}
import replicatn as rtn
B   = rtn.SymmetricBasis(2)
d   = 2
bd  = rtn.IPRBoundary(B, d)        # the XEB boundary is the same as IPR
chans_clean = [rtn.identity_choi(d),     rtn.identity_choi(d)]
chans_dep   = [rtn.identity_choi(d),     rtn.depolarising_choi(d, p)]
val = rtn.noisy_brickwork_average(B, d, N, t, bd, chans_dep)
\end{pybox}
As a simple test, we will use \emph{incoherent} depolarising noise at rate $p$.

Fig.~\ref{fig:xeb_k2} reports $\chi_{\mathrm{XEB}}(t)$ on $N \in \{8, 16, 32, 64\}$ qubits up to depth $t = 40$, in three panels: (a) clean (no device error), (b) depolarising noise on the device replica at rate $p = 0.10$, and (c) the same at $p = 0.20$.  
The clean curve in (a) recovers the Haar-fidelity plateau $\chi_{\mathrm{XEB}}(t) \to 2 d^N / (d^N + 1) - 1 = (D - 1) / (D + 1) \to 1$ at depths $t \gtrsim \log N$, as expected from substituting $d^N\,\mathbb{E}[\sum_x p^2] = 2 d^N / (d^N + 1)$ into Eq.~\eqref{eq:xeb_def}.  In (b) and (c) the device noise drives $\chi_{\mathrm{XEB}}$ exponentially down to its classical (no-quantum-advantage) plateau $\chi_{\mathrm{XEB}} = 0$. 

tracks the running fidelity of the noisy circuit relative to its ideal simulation, which is the operational content of the XEB benchmark.

\begin{exercise}
\label{sec:exercise_setupII_finite_rate}
\textbf{Finite-rate error-correction transition.}
Fig.~\ref{fig:coherent_info} uses $K = 1$ logical qubit, which gives a
smooth crossover in $t$ but no \emph{sharp} finite-size effect.  Following Ref.~\cite{sauliere2026errcorr}, lift the construction to a finite encoding rate $r = K / N_B$ held constant
as $N_B$ grows, i.e.\ $K = r\,N_B$ Bell pairs between $R$ and the first
$r\,N_B$ sites of $B$.  At fixed $r > 0$ the curves $I_c^{(2)} / k$ at
different $(N_B, t)$ collapse onto a universal function of the
cumulative-noise parameter $f_2 \propto p\,t$ and develop a sharp
critical point at $f_2^{\rm crit} = (1 - r)(1 + g_{s,e})$, where
$g_{s,e} = \log \tilde G_{s, e}(d, \mathcal{N})$ encodes the
depolarising channel.  
(a) Build the initial state, final state and bulk circuit for the RTN at a given input $K$. 
(b) For $r = 1/4$ and a small grid of
$(N_B, p)$, sweep $t$ and record $I_c^{(2)} / k$. 
(c) Obtain the function $F = \mathbb{E}_U \!\left[ \mathrm{tr}\!\left[ \mathcal{U}_t\!\bigl(\ket{0}\!\bra{0}\bigr)\, U\ket{0}\!\bra{0}U^\dagger \right] \right]$ with the appropriate RTN boundary state, and compute $f_2:=-\tfrac{2}{N}\log \tilde{F}$ with $\tilde{F}=F-d^{-N}$.
(d) Reproduce the phase transition diagram plotting $I_c/K$ against $f_2$. 
\end{exercise}

\section{Random circuits beyond Haar unitary}
\label{sec:beyond_haar_unitary}

The statistical mechanics mapping of \S\S\ref{sec:rtn:superoperator}--\ref{sec:rtn:mps_evolution} extends naturally to other gate ensembles.  The key abstraction is the \emph{commutant}: given an ensemble $\mathcal{E}$ of $q\times q$ unitaries, the operators that commute with $k$ pairs $U \otimes U^*$ for every $U \sim \mathcal{E}$ form a finite-dimensional algebra $\mathrm{Comm}_k(\mathcal{E})$.  Picking a basis $\{|\sigma\rrangle : \sigma \in \mathcal{B}_k(\mathcal{E})\}$ of this commutant and lifting it to the doubled $k$-replica space, we recover the same RTN construction with $\mathrm{S}_k$-spins replaced by spins of dimension $q_\mathrm{eff} = |\mathrm{Comm}_k(\mathcal{E})|$, the $\mathrm{S}_k$ Gram matrix of~Eq.~\eqref{eq:gram_def} replaced by the analogous $\mathcal{E}$-Gram
\begin{equation}
G^{\mathcal{E}}_{\sigma, \tau}(q) \;=\; \llangle \sigma|\tau\rrangle\bigr|_{\dim = q} ,
\qquad
\mathrm{Wg}^{\mathcal{E}}(q) = \bigl(G^{\mathcal{E}}(q)\bigr)^{+} ,
\label{eq:generic_gram}
\end{equation}
and the Haar moment formula~Eq.~\eqref{eq:haar_moment} replaced by the universal expansion
\begin{equation}
\mathbb{E}_{U \sim \mathcal{E}}\bigl[(U \otimes U^*)^{\otimes k}\bigr]
\;=\;
\sum_{\sigma, \pi \in \mathcal{B}_k(\mathcal{E})}
   \mathrm{Wg}^{\mathcal{E}}_{\sigma, \pi}(d^2)\,
   |\sigma\rrangle\llangle\pi| .
\label{eq:moment_operator_general}
\end{equation}
The rest of the machinery, dressed transfer matrix, MPS evolution, boundary contraction, is identical to the unitary-Haar case.  Implementing a new ensemble therefore reduces to (i) identifying the commutant basis $\mathcal{B}_k(\mathcal{E})$ and (ii) writing the closed-form $\llangle \sigma|\tau\rrangle$ in terms of its combinatorial data; the companion package abstracts both behind a single \texttt{CommutantBasis} interface.  In the rest of this section we work out two concrete cases: the orthogonal group $\mathrm{O}(q)$ (\S\ref{sec:beyond_haar:orthogonal}) and the Clifford group on qutrits (\S\ref{sec:beyond_haar:clifford}).

\smallskip\noindent\textit{A note on notation.}  In what follows we
keep the symbol $|\sigma\rrangle$ for any element of the commutant
basis $\mathcal{B}_k(\mathcal{E})$, even when $\sigma$ is no longer a
permutation: for the orthogonal group $|\sigma\rrangle$ stands for a
Brauer diagram (perfect matching), while for the Clifford ensemble it stands
for a stochastic Lagrangian subspace.  This abuse of notation is
deliberate, it lets the same MPS-contraction code, with the same
boundary kets, run unchanged on every ensemble.

\subsection{Orthogonal Haar circuits}
\label{sec:beyond_haar:orthogonal}

\label{sec:orthogonal}

For two-qudit gates drawn from the orthogonal group $\mathrm{O}(d^2)$ instead of $\mathrm{U}(d^2)$, the commutant of $\mathrm{O}(q)$ acting diagonally on $(\mathbb{C}^q)^{\otimes k}$ is the \emph{Brauer algebra} $B_k(q)$~\cite{brauer1937algebras}.  Its natural basis consists of the perfect matchings of $2k$ labelled points: $k$ ``upper'' (ket) and $k$ ``lower'' (bra) vertices, paired in any way that produces a fixed-point-free involution.  In addition to the propagating strands, which reproduce the symmetric group $\mathrm{S}_k \subset B_k$, generic Brauer diagrams contain \emph{caps} (two upper vertices paired) and \emph{cups} (two lower vertices paired); the total number is $|B_k| = (2k - 1)!! = 1, 3, 15, 105, 945, \ldots$ for $k = 1, 2, 3, 4, 5$.  The Brauer Gram matrix is given by the loop-counting formula
\begin{equation}
G^{\mathrm{O}}_{m, m'}(q) \;=\; q^{L(m, m')} ,
\label{eq:brauer_gram}
\end{equation}
where $L(m, m')$ is the number of disjoint cycles produced by stacking $m$ on top of $m'$ and identifying the middle vertices.

\begin{figure}[t!]
\centering
\includegraphics[width=0.7\textwidth]{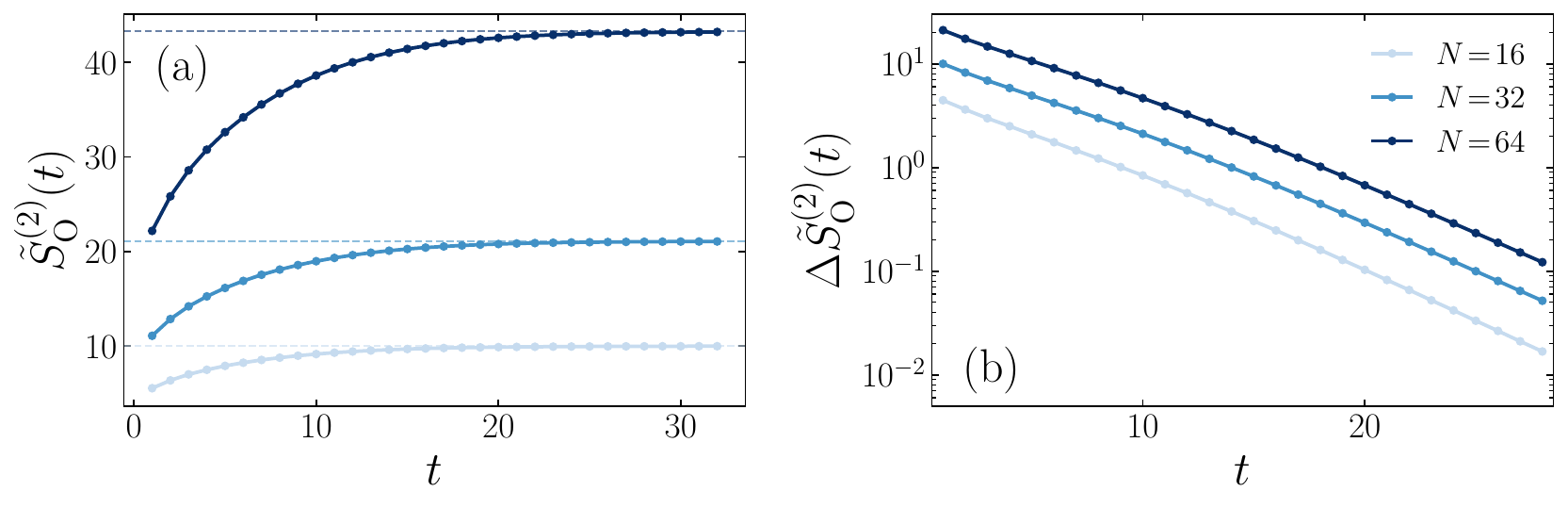}
\caption{(a) $\tilde{S}^{(2)}_O:=-\log \mathbb{E}[\mathcal{I}^{(2)}_t]$ for an orthogonal-Haar brickwork on qubits, plotted as a function of circuit depth~$t$ for $N \in \{8, 16, 32, 64\}$.  The horizontal asymptote is the orthogonal stationary value~Eq.~\eqref{eq:ipr_orthogonal_stat}. (b) Difference $\Delta \tilde{S}^{(2)}_O(t)=\tilde{S}^{(2)}_O(\infty) - \tilde{S}^{(2)}_O(t)$ is plotted for different system sizes, highlighting saturation at fixed tolerance $\varepsilon$ in a timescale that is logarithmic in system size $t_\mathrm{AC}\simeq \log(N)$.}
\label{fig:ipr_orthogonal}
\end{figure}

At $k = 2$, the Brauer algebra has three elements
$|B_2| = (2 \cdot 2 - 1)!! = 3$: the identity $\iota$, the swap $s$ already familiar from the symmetric group, and one additional element, the \emph{crossing} diagram, which we denote $|c\rrangle$.  Concretely, $|c\rrangle = \sum_{a, b}|a, a, b, b\rrangle$ glues the two upper vertices into a cap and the two lower vertices into a cup, producing the contraction $\sum_a |a, a\rangle\langle a, a|$ that is invariant under $\mathrm{O}(q)$ but not under $\mathrm{U}(q)$.  The IPR boundary $|\Omega^{\textup{IPR}}_2\rrangle$ of~Eq.~\eqref{eq:boundary_IPR} is unchanged (the computational-basis constraint does not see the choice of ensemble), and its overlap with $|c\rrangle$ also gives $d$.  The orthogonal IPR therefore admits the closed-form stationary value (single global gate)
\begin{equation}
\mathbb{E}_{\mathrm{O}}\bigl[\mathcal{I}^{(2)}\bigr]
\;=\;
\frac{3}{D + 2} ,
\qquad D = d^N ,
\label{eq:ipr_orthogonal_stat}
\end{equation}
where the numerator $3 = (2 \cdot 2 - 1)!! = 3!!$ is the orthogonal-Haar
double factorial.  The general $k$th moment on $\mathrm{O}(D)$ reads
$\mathbb{E}_{\mathrm{O}}[\mathcal{I}^{(k)}] = (2k - 1)!! / [(D + 2)(D + 4)\cdots(D + 2k - 2)]$,
and for $k = 2$ it differs from the unitary value $2/(D + 1)$ by the
extra ``crossing'' diagrams of $\mathrm{O}(D)$~\cite{sauliere}.  Plugging the Brauer Gram and Weingarten into the same dressed transfer matrix as the unitary case yields the local IPR dynamics; in code, switching from unitary to orthogonal is the one-liner
\begin{pybox}
import replicatn as rtn
B_O = rtn.BrauerBasis(2)              # |B_O| = 3 (orthogonal, k = 2)
val = rtn.brickwork_average(B_O, d, N, t, rtn.IPRBoundary(B_O, d))
\end{pybox}

The corresponding $\tilde{S}^{(2)}_O:=-\log \mathbb{E}[\mathcal{I}^{(2)}_t]$ is plotted in Fig.~\ref{fig:ipr_orthogonal}(a). As for the unitary case, the saturation at a fixed precision to the orthogonal Haar value Eq.~\eqref{eq:ipr_orthogonal_stat} occurs in a timescale that is logarithmic in system size.

\begin{exercise}
\label{sec:exercise_otoc_orthogonal}
\textbf{OTOC for orthogonal-Haar circuits.}
The OTOC carries over verbatim to orthogonal-Haar: same boundary recipe, same MPS sweep, only the commutant changes.  At $k = 2$ the Brauer algebra has $|\mathcal{B}_2| = 3$ basis elements (id, swap, and the crossing) and the OTOC profile is the column $G^O(s,\cdot;d)$ of the \emph{orthogonal} Gram~Eq.~\eqref{eq:brauer_gram}. 
(a) Compute the stationary (orthogonal Haar) value for $A_1=A_2=B_1=B_2=Z_1$.
(b) Implement the dynamics of the OTOC for a brickwork orthogonal circuit. At what timescale saturates to the Haar value? How does it behave compared to the unitary case?
\end{exercise}

\subsection{Clifford random circuits}
\label{sec:beyond_haar:clifford}

\label{sec:applications_3}
\label{sec:hands_on_3}

When the two-qudit gates are drawn uniformly from the Clifford group $\mathcal{C}_N$ rather than from the full unitary group, the commutant for general $k$ is generated by the \emph{stochastic Lagrangian subspaces} of $\mathbb{F}_d^{2k}$~\cite{gross2021schurweyl}.  At $k = 2$ the Clifford ensemble is a unitary $2$-design for every $d$, so the Clifford and Haar moments coincide and~Eq.~\eqref{eq:moment_operator_general} reduces to the standard $\mathrm{S}_2$ formula. 
For $d>2$, the first non-trivial case is at $k = 3$.  To make the discussion concrete we take qutrits ($d = 3$), where one already feels the difference between the Clifford and unitary moments while the bookkeeping stays small.

\begin{figure}[t!]
\centering
\includegraphics[width=0.7\textwidth]{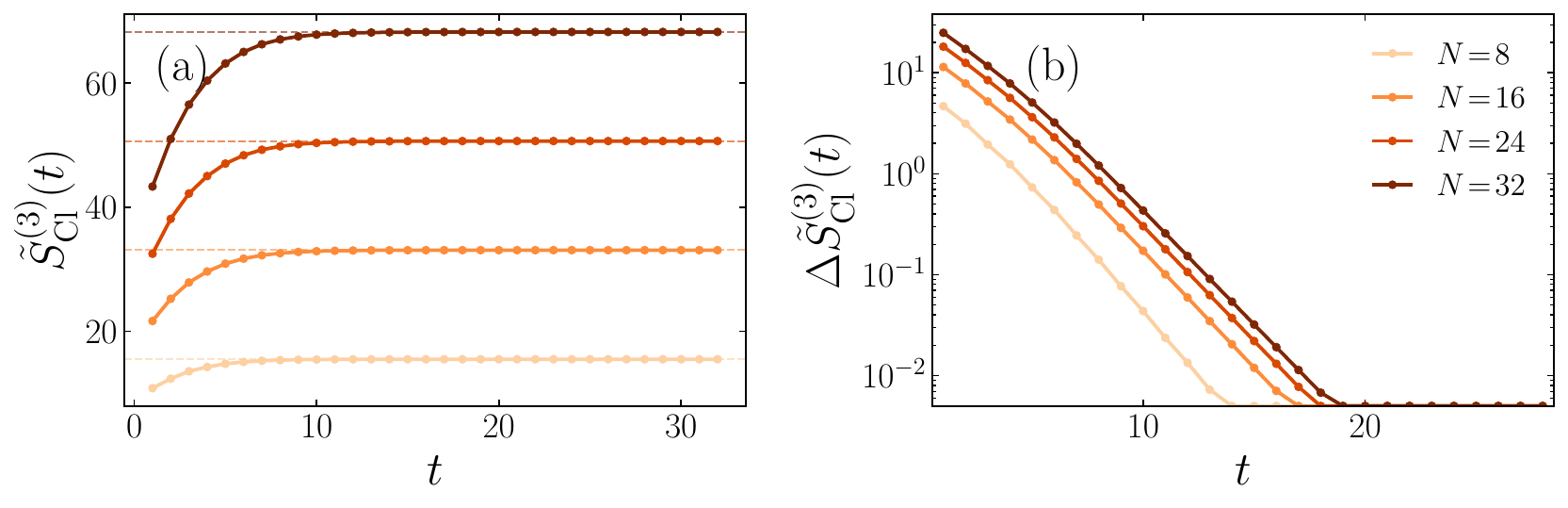}
\caption{(a) $\tilde{S}^{(3)}_\mathrm{Cl}:=-\tfrac{1}{2}\log \mathbb{E}[\mathcal{I}^{(3)}]$ on qutrits for a Clifford-random brickwork (markers). (b) The difference $\Delta\tilde{S}^{(3)}_\mathrm{Cl}(t)$ is plotted for various system sizes. }
\label{fig:ipr_clifford}
\end{figure}

For qutrits at $k = 3$, the Clifford commutant has $|\mathcal{B}_3^{\mathrm{Cl}}| = 8$ basis elements, two more than the symmetric-group commutant $|\mathrm{S}_3| = 6$.  For the $i$th replica qudit, the two extra elements are the state
\begin{equation}
\begin{split}
|Q_3()\rrangle_i \;&=\; \left(\frac{1}{d}\,\sum_{P \in \mathcal{P}_d}\,(I\otimes P)^{\otimes 3}\right) |\iota\rrangle_i,\\
|Q_3(12)\rrangle_i \;&=\; \left(\frac{1}{d}\,\sum_{P \in \mathcal{P}_d}\,(I\otimes P)^{\otimes 3}\right) |s\rrangle_i,
\label{eq:Q3_clifford}
\end{split}
\end{equation}
where $s=(12)$ is the swap acting on replicas $1$ and $2$, and $\mathcal{P}_d = \{P_a\}_{a=1}^{d^2}$ is an orthonormal Pauli basis on a single qutrit. 
The total replica states are given by $|Q_3 \sigma\rrangle=\bigotimes_{i=1}^N |Q_3\sigma\rrangle$ for $\sigma=(), (12)$. 
Plugging the eight-element commutant into~Eq.~\eqref{eq:moment_operator_general} and contracting with the IPR boundary gives the closed-form Clifford stationary value of the third IPR.  Eq.~(20) of~\cite{magni2025anticoncentration} (see also~\cite{christopoulos2024universal}) writes it as the ratio of two $q$-Pochhammer symbols,
\begin{equation}
\mathbb{E}_{\mathrm{Cl}}\bigl[\mathcal{I}^{(k)}\bigr]
\;=\;
\frac{(-d^{\,2-k};\,d)_N}{(-d;\,d)_N} ,
\qquad
(a;\,\xi)_N \;=\; \prod_{m=0}^{N-1}\bigl(1 - a\,\xi^m\bigr) ,
\label{eq:ipr_clifford_stat}
\end{equation}
which on qutrits at $k = 3$ gives $I_\mathcal{C}^{(3)}:=8/[(D+1)(D+3)]$.  
To study the IPR dynamics, in code the only change with respect to the unitary brickwork is to swap the commutant basis,
\begin{pybox}
import replicatn as rtn
B_Cl = rtn.CliffordBasis(3, 3)        # |B_Cl| = 8 (Clifford, qutrits, k = 3)
val  = rtn.brickwork_average(B_Cl, 3, N, t, rtn.IPRBoundary(B_Cl, 3))
\end{pybox}
which runs the same MPS sweep as in the unitary case on the larger commutant.  Fig.~\ref{fig:ipr_clifford} reports the dynamics of $\tilde{S}^{(3)}_\mathrm{Cl}:=-\tfrac{1}{2}\log \mathbb{E}[\mathcal{I}^{(3)}]$ on qutrits for the Clifford ensemble: panel (a) shows the bare RTN data converging to the dotted Clifford plateau, cf. Eq.~\eqref{eq:ipr_clifford_stat}, while panel (b) plots $\Delta \tilde{S}^{(3)}_\mathrm{Cl}$ the difference between the stationary state and the time-evolving one. 

\FloatBarrier
\section{Discussion}
\label{sec:discussion}

The replica tensor network reduces the calculation of any
one-dimensional circuit-averaged polynomial of degree $k$ in the gates to
the contraction of a two-dimensional statistical-mechanics model on the
commutant of the underlying gate group.  The numerical cost is governed
by three parameters only --- the replica count $k$, the local qudit
dimension $d$, and the effective MPS bond dimension $O(|\mathcal{B}|^2)$
--- and is essentially independent of $d^N$.  This makes the RTN a
practical device for hundreds of qudits, well beyond the reach of direct
state-vector simulation, while preserving analytical control through the
commutant-basis representation.

For simplicity we covered only the simplest scenarios, yet still tackled
physically motivated quantities: inverse participation ratios and
participation entropies, and measures of error-resilience in many-body
systems.  The strength of the method is rooted in its versatility:
changing the gate, or adding noise or deterministic injections, only
modifies the bulk tensor via generalised Weingarten calculus; changing
observables or initial states only modifies the replica boundary
conditions.

Several directions not covered in these notes are immediate extensions
of the same machinery.  On the geometric side, while we focused on
brickwork circuits, alternative geometries adapt naturally: random
matrix product states and random tensor networks are two such
classes~\cite{lami2024anticoncentration}, as is the gluing of quantum
circuits~\cite{schuster_random_2025,magni2025anticoncentration}.

When the local Hilbert-space dimension is sufficiently large, a further simplification can be obtained using free-probability methods~\cite{PhysRevLett.129.170603,PhysRevX.15.011031,PhysRevLett.134.140404,dowling2025freeindependenceunitarydesign,zhang2026finitesizescalingeigenstatethermalization,fritzsch2025freeprobabilityminimalquantum,fritzsch2025freecumulantseigenstatethermalization,herrmann2026timescalesdeepthermalization,vallini2025refinementseigenstatethermalizationhypothesis,prv4-948b,dowling2026pagecurvelocaloperatorentanglement}. This reduces the number of internal degrees of freedom from $|\mathrm{S}_k|$ to the Catalan number $C_k$, providing a complementary route to the irreducible-representation reduction for qubits discussed in Sec.~\ref{sec:irrep_reduction}.

On the symmetry side, enforcing a global $U(1)$
conservation~\cite{rakovszky_diffusive_2019,turkeshi2024quantumMpemba,aditya},
or more general finite-group symmetries, refines the commutant basis
into charge-resolved blocks.  Other ensembles, such as fermionic
circuits, can be handled by the same machinery, building on recent
advances in the matchgate
commutant~\cite{sierant2026theorymatchgatecommutant,braccia2026commutantfermionicgaussianunitaries,lastres2026geometryfreefermioncommutants}. 
Even measurements can be added, provided additional care is taken on the replica extrapolations, cf. Ref.~\cite{wlj6-mkk4,gerbino2026universalpurificationdynamicsreal,lami2025quantumstatedesignemergent}
Each of these extensions reuses the same two abstractions, a commutant-basis spec and a boundary ket, that we have introduced here, and we hope that the present notes serve as a stepping stone for the reader to take them up.

\section*{Acknowledgements}
I thank especially Piotr Sierant for extensive collaborations and discussions on the subjects related to these notes; Beatrice Magni, Sreemayee Aditya, Riccardo Cioli, Arman Sauliere, Guglielmo Lami, Andrea De Luca, and Jacopo De Nardis for collaborations on related topics, and Adam Nahum for discussions. 

This work is supported by DFG under:  Germany's Excellence
Strategy -- Cluster of Excellence Matter and Light for Quantum
Computing (ML4Q) EXC 2004/2 -- 390534769,  the Collaborative
Research Center (CRC) 183 Project No.~277101999 -- project~B01,
and Emmy Noether Programme proposal ``Digital Quantum Matter
Out-of-Equilibrium'' No.~560726973.

This work is dedicated to Ada. 

\textbf{Code and Data Availability.}
The companion package \texttt{ReplicaTN} (v0.3.0) is
released under the MIT license at
\url{https://github.com/xturkesh/ReplicaTN}, to appear in Zenodo~\cite{xhekzenodo} in a future version with the data.

\begin{appendix}
\numberwithin{equation}{section}

\FloatBarrier
\section{Numerical implementation walk-through}
\label{sec:implementation_observables}

\subsection{Introduction to the numerical implementation}

We established that the circuit-averaged quantities can be computed as the contraction of a two-dimensional tensor network with $k!$-state effective spins.  
While I encourage the reader to reproduce with their own code the ideas discussed here, e.g. supported by open libraries such as \texttt{ITensor} or \texttt{TeNPy}~\cite{itensor,pollo}, I attach with the notes a minimalistic C++ codebase with Python frontend via \texttt{pybind11} that build the Gram matrix, the averaged gate tensor, and the brickwork MPS contraction directly from the formulae of Sec.~\ref{sec:rtn_setup}. 
Every call we use below is one of \texttt{SymmetricBasis}, \texttt{gram\_matrix}, \texttt{IPRBoundary}, \texttt{brickwork\_rtn\_average} (and their noisy/orthogonal/Clifford analogues).

The full workflow is summarised in Algorithm~\ref{alg:rtn_brickwork}.  The strategy it encodes is:
\begin{enumerate}
\item Build the local tensors: the Gram matrix, the Weingarten matrix, and the replica transfer matrix.
\item Encode the initial state and boundary conditions as an MPS.
\item Evolve the MPS through each layer of the replica tensor network using the transfer matrix as an MPO.
\item Contract with the final boundary to extract the result.
\end{enumerate}
We focus on the collision probability ($k = 2$) for concreteness, where the effective spin dimension is $q_{\mathrm{eff}} = 2! = 2$, but the code generalises to arbitrary $k$ and other replica boundary conditions.

\begin{Algorithm}{Replica tensor-network contraction of a brickwork circuit average.\label{alg:rtn_brickwork}\vspace{-0.2cm}}
\Require commutant basis $\mathcal{B}_k$ of size $n_\mathcal{B}$; qudit dimension $d$; chain length $N$; depth $t$; boundary vector $b \in \mathbb{R}^{n_\mathcal{B}}$ (or site-dependent profile) encoding the observable.
\State $G \gets$ $n_\mathcal{B} \times n_\mathcal{B}$ Gram matrix, $G_{\sigma,\tau} = \llangle \sigma | \tau \rrangle(d)$
\State $G_{(2)} \gets $ Gram matrix at $q = d^2$
\State $\mathrm{Wg}_{(2)} \gets G_{(2)}^{-1}$  \Comment{pseudo-inverse when $d^2 < k$, else the exact inverse}
\State build the 4-index averaged gate $T[\sigma_1, \sigma_2, \tau_1, \tau_2] = \delta_{\sigma_1, \sigma_2} \sum_\pi \mathrm{Wg}_{(2), \sigma_1, \pi} \, G_{\pi, \tau_1} \, G_{\pi, \tau_2}$
\State set the per-site initial amplitude $c_\mathrm{init}(\sigma) = \sum_\tau \mathrm{Wg}(\sigma, \tau; d)$  \Comment{here for the product state $|0\rangle^{\otimes N}$}
\State build the product MPS $|\psi_0\rangle = \bigotimes_{i=1}^{N} c_\mathrm{init}$
\For{$\ell = 1$ \textbf{to} $t$}
  \State $\mathcal{P} \gets \{(1,2), (3,4), \dots\}$ if $\ell$ odd, else $\{(2,3), (4,5), \dots\}$  \Comment{brickwork pattern}
  \For{$(i, i+1) \in \mathcal{P}$}
    \State apply $T$ on sites $(i, i+1)$, SVD-truncate to bond dim $\chi$
  \EndFor
\EndFor
\State $\langle \Lambda \rangle \gets \langle b^{\otimes N} | \psi_t \rangle$  \Comment{inner product with the observable boundary}
\State \Return $\langle \Lambda \rangle$
\vspace{-0.2cm}
\end{Algorithm}

\subsection{Permutation basis, Gram and Weingarten matrices}

The effective spins of the replica tensor network are elements of
the commutant basis returned by \texttt{SymmetricBasis(k)} (size
$|\mathrm{S}_k| = k!$ for the unitary case; orthogonal and Clifford analogues
swap in \texttt{BrauerBasis(k)} or \texttt{CliffordBasis(k, d)}).
The two algebraic ingredients of the contraction are the Gram matrix
$G_{\sigma, \tau}(\chi) = \chi^{\#(\sigma^{-1}\tau)}$ of
Eq.~\eqref{eq:gram_def} and its Moore--Penrose pseudo-inverse
$\mathrm{Wg}(\chi) = G(\chi)^{+}$ of
\S\ref{subsec:weingarten}, both tabulated at the two
dimensions that appear in the brickwork: $\chi = d$ on a single site,
and $\chi = d^2$ on a two-site averaged gate.  In code,

\begin{pybox}
import replicatn as rtn
import numpy as np

d = 2                                   # qubits
k = 2                                   # collision probability
B   = rtn.SymmetricBasis(k)               # |B| = k! for unitary
G   = rtn.gram_matrix(B, d)               # |B| x |B| at single-site dim d
G2  = rtn.gram_matrix(B, d * d)           # at two-site dim d^2
Wg2 = rtn.weingarten_matrix(B, d * d)     # = inv(G2)
assert np.allclose(G2 @ Wg2, np.eye(len(B)))
\end{pybox}
\noindent The pseudo-inverse \texttt{pinv} matters only when
$\chi < k$, where $G(\chi)$ has a non-trivial kernel because the
permutation states become linearly dependent in the doubled
single-site space (see~\cite{collins_weingarten_2022} and
Appendix~\ref{app:permutations}); for the brickwork at $\chi = d^2$
and $k \leq 3$ the matrix is invertible and \texttt{pinv} returns
the exact inverse.

\subsection{Building the transfer matrix as an MPO}

The central object is the replica transfer matrix $\mathcal{T}^{(k)}_{i,i+1}$, which acts on two neighbouring replica spins. For $k = 2$, this is a $4 \times 4$ matrix (input: two $q_{\mathrm{eff}}$-dimensional spins; output: two $q_{\mathrm{eff}}$-dimensional spins).

The dressed transfer matrix~Eq.~\eqref{eq:transfer_dressed} is constructed as follows:
\begin{enumerate}
\item Compute the two-site Weingarten matrix: $\mathrm{Wg}(d^2)$.
\item Compute the single-site Gram matrix: $G(d)$.
\item The tensor element $\widetilde{\mathcal{T}}^{\sigma_1, \sigma_2}_{\tau_1, \tau_2}$ couples input permutations $(\tau_1, \tau_2)$ to output permutations $(\sigma_1, \sigma_2)$ via
\begin{equation}
\widetilde{\mathcal{T}}^{\sigma_1, \sigma_2}_{\tau_1, \tau_2} = \sum_{\pi, \pi' \in \mathrm{S}_k} \mathrm{Wg}_{\sigma \cdot \sigma', \pi \cdot \pi'}(d^2) \cdot G_{\pi, \tau_1}(d) \cdot G_{\pi', \tau_2}(d),
\end{equation}
where $\sigma \cdot \sigma'$ denotes the combined two-site permutation.
\end{enumerate}

\begin{pybox}
T_gate = rtn.averaged_gate_tensor(B, d)
print(T_gate.shape)            # ( |B|, |B|, |B|, |B| )  i.e. (2,2,2,2) for k=2
\end{pybox}

Internally, \texttt{averaged\_gate\_tensor} builds the same four-index object as~Eq.~\eqref{eq:transfer_dressed} from the Gram and Weingarten matrices already returned by \texttt{gram\_matrix(B, d)} and by the fucntion \texttt{weingarten\_matrix(B, d\^{}2)}. The result is a plain $|B|^2 \times |B|^2$ \texttt{numpy} array that the brickwork driver below reshapes into a rank-4 tensor on demand.

For $k=2$ and $d=2$, the effective local dimension is $|B| = 2$, so the transfer tensor has $2^4 = 16$ entries.  For $k = 3$ ($|B| = 6$) or $k = 4$ ($|B| = 24$) the tensor grows to $36^2$ and $576^2$ entries, respectively (however, these tensors are very sparse and most entries are zeros). The brickwork contraction proceeds at MPS bond dimension $\chi \sim O(|B|^2)$, which stays modest at the values of $k$ we use in practice.

\subsection{Setting up the MPS and evolving}

We now describe the full pipeline for computing the average IPR via MPS evolution.

\begin{pybox}
def compute_avg_ipr(N: int, t: int, d: int, k: int, chi_max: int = 100):
    """Circuit-averaged kth IPR of a depth-t Haar brickwork on N qudits of
    dimension d, computed via the high-level ReplicaTN driver."""
    import replicatn as rtn
    B  = rtn.SymmetricBasis(k)
    bd = rtn.IPRBoundary(B, d)
    return rtn.brickwork_average(B, d, N, t, bd,
                                cutoff=1e-13, maxdim=chi_max)
\end{pybox}

The driver \texttt{brickwork\_rtn\_average} takes care of (i) building the boundary product MPS \texttt{init\_amplitudes(B, d)} on every site, (ii) applying \texttt{averaged\_gate\_tensor(B, d)} on the odd / even bonds in turn, (iii) SVD-truncating to bond dimension \texttt{chi\_max}, (iv) contracting against the per-site IPR boundary $b(\sigma) = d$.  All four steps go through the minimal \texttt{ReplicaTN} tensor network library (\texttt{cpp/ReplicaTN.h}), cf. Ref.~\cite{itensor,pollo} for similar (and more versatile!) implementations. The Python frontend (\texttt{python/replicatn/}) wraps them via \texttt{pybind11} and adds the symbolic algebra (commutants, Weingarten, gates, boundaries) on top in \texttt{algebra.py}.

\subsection{A concrete example: collision probability for $N$ qubits}

Let us put the pieces together and compute the collision probability $\mathcal{I}^{(2)}$ for $N$ qubits as a function of the circuit depth $t$.

\begin{pybox}
from math import factorial

d   = 2          # qubits
k   = 2          # collision probability
chi = 50         # bond dimension (more than enough for k = 2)
Ns    = [16, 32, 64, 128, 256, 512]
t_max = 20

def ipr_haar(D, k):
    return factorial(k) * factorial(D - 1) / factorial(D + k - 1)

for N in Ns:
    D = d ** N
    ipr_H = ipr_haar(D, k)
    print(f"N = {N}, IPR_Haar = {ipr_H:.4e}")
    for t in range(1, t_max + 1):
        ipr_avg = compute_avg_ipr(N, t, d, k, chi_max=chi)
        delta   = ipr_avg - ipr_H
        print(f"  t = {t}: IPR = {ipr_avg:.4e}, Delta = {delta:.4e}")
\end{pybox}

\textit{Convergence with bond dimension.}For $k=2$, the MPS bond dimension required for convergence is modest: $\chi \sim q_{\mathrm{eff}}^2 = 4$ is often sufficient. For $k=3$ ($q_{\mathrm{eff}} = 6$), one needs $\chi \sim 36$, and for $k=4$ ($q_{\mathrm{eff}} = 24$), $\chi \sim 576$. These values reflect the fact that the entanglement in the replica tensor network saturates at a level determined by the effective local dimension~\cite{turkeshi2025magic}.

\end{appendix}

%
\end{document}